\documentclass[apj]{emulateapj}

\usepackage{epsfig}
\usepackage{amssymb}
\usepackage{amsmath}
\usepackage{rotating}

\def\apj{{\it ApJ \,}}
\def\apjs{{\it ApJS} \,}

\def\apjl{{\it ApJ} \,}

\def\prd{{\it Phys. Rev. D.} \,}
\def\mnras{{\it MNRAS} \,}

\def\na{{\it NewA} \,}

\def\aap{{\it A\&A} \,}

\def\jcap{{\it JCAP} \,}
              
\begin{document}

\title[Dark Matter Velocity Anisotropy in Clusters]{Profiles of Dark Matter 
Velocity Anisotropy in Simulated Clusters} 

\author{
Doron Lemze\altaffilmark{1},
Rick Wagner\altaffilmark{2},
Yoel Rephaeli\altaffilmark{3,4},
Sharon Sadeh \altaffilmark{4},
Michael L. Norman\altaffilmark{2,3,5},
Rennan Barkana\altaffilmark{4}, \\
Tom Broadhurst\altaffilmark{6,7},
Holland Ford \altaffilmark{1},
\& Marc Postman \altaffilmark{8}
}
\altaffiltext{1}
                {Department of Physics and Astronomy, Johns Hopkins University,
3400 North Charles Street, Baltimore, MD 21218, USA; doronl@pha.jhu.edu}

\altaffiltext{2}
               {San Diego Supercomputer Center, University of California, San
Diego, MC 0505, 10100 Hopkins Drive, La Jolla, CA 92093, USA}

\altaffiltext{3}
                {Center for Astrophysics and Space Sciences, University of
California, San Diego, La Jolla, CA 92093, USA}

\altaffiltext{4}
                {School of Physics and Astronomy, Tel Aviv University, Tel Aviv,
69978, Israel}

\altaffiltext{5} {Physics Department, University of California, San Diego, La
Jolla, CA 92093, USA}

\altaffiltext{6}
                {Department of Theoretical Physics, University of Basque Country
UPV/EHU,Leioa,Spain}

\altaffiltext{7}
                {IKERBASQUE, Basque Foundation for Science,48011, Bilbao,Spain}

\altaffiltext{8}
                {Space Telescope Science Institute, Baltimore, MD 21218, USA}

\begin{abstract}

We report statistical results for dark matter (DM) velocity anisotropy, $\beta$,
from a sample of some 6000 cluster-size halos (at redshift zero) identified in a
$\Lambda$CDM hydrodynamical adaptive mesh refinement simulation performed
with the {\it Enzo} code. These include profiles of $\beta$ in clusters with
different masses, relaxation states, and at several redshifts, modeled both as
spherical and triaxial DM configurations. Specifically, although we find a
large scatter in the DM velocity anisotropy profiles of different halos (across
elliptical shells extending to at least $\sim 1.5 r_{\rm vir}$), universal
patterns are found when these are averaged over halo mass, redshift, and
relaxation stage. These are characterized by a very small velocity anisotropy at
the halo center, increasing outward to $\sim 0.27$ and leveling off at $\sim
0.2 r_{\rm vir}$. Indirect measurements of the DM velocity anisotropy fall on
the upper end of the theoretically expected range. Though measured indirectly,
the estimations are derived by using two different surrogate measurements -
X-ray and galaxy dynamics. Current estimates of the DM velocity anisotropy are
based on very small cluster sample. Increasing this sample will allow testing
theoretical predictions, including the speculation that the decay of DM
particles results in a large velocity boost. We also find, in accord with
previous works, that halos are triaxial and likely to be more prolate when
unrelaxed, whereas relaxed halos are more likely to be oblate. Our analysis does
not indicate that there is significant correlation (found in some previous
studies) between the radial density slope, $\gamma$, and $\beta$ at large radii,
$0.3\;r_{\rm vir} < r < r_{\rm vir}$.

\end{abstract}

\keywords{Methods: Numerical -- Galaxies: clusters: general}

\section{Introduction}
\label{Introduction}

Dark matter (DM), the main mass constituent of galaxy clusters, dominates 
the dynamics of intracluster (IC) gas and member galaxies. The DM mass density 
profile was until recently the only cluster property that could be inferred 
from simulations and tested against observational data. The DM velocity
anisotropy profile also holds important information, since it depends on both
the DM particle features, such as its collisionless nature (e.g. Host et al.\
2009) and decaying time (Peter, Moody, \& Kamionkowski 2010), and the wide
variety of dynamical collapse processes (Wang \& White 2009). In addition, halo
mass dynamical estimators, such as Jeans (Binney \& Tremaine 2008) and the
caustics (Diaferio \& Geller 97; Diaferio 99), that use galaxy information,
depend on the galaxy velocity anisotropy. Since large hydrodynamical
cosmological simulations that include galaxies are not found, one can use the
DM velocity anisotropy, since both (DM and galaxies) are considered to be
collisionless, and therefore should have similar dynamical properties. That
said, measured galaxy velocity anisotropy profiles (e.g. Benatov et al.\ 2006;
Lemze et al.\ 2009) maybe somewhat different than the ones estimated for DM
using simulations. Some effort is now devoted to determine also the DM velocity
anisotropy either by using the gas temperature as a tracer of the DM velocity
anisotropy, a method which is applicable at intermediate radii (Host et al.\
2009), or by examining galaxy velocities, as has recently been demonstrated in
the analysis of A1689 measurements (Lemze et al.\ 2011). It is in fact our 
plan to apply the latter procedure to additional relaxed X-ray clusters 
in the CLASH program (Postman et al.\ 2012).

N-body simulations (for various cosmological models) suggest a nearly
universal velocity anisotropy profile (Cole \& Lacey 1996; Carlberg et al.\
1997; Colin, Klypin, \& Kravtsov 2000; Diemand, Moore, \& Stadel\ 2004; Rasia et
al.\ 2004; Wojtak et al.\ 2005), similar to the universal DM density
profile deduced from simulations (Navarro, Frenk, \& White 1997, hereafter NFW;
Moore et al.\ 1998)and various observations (X-ray: e.g. Pointecouteau, Arnaud,
\& Pratt 2005; Arnaud, Pointecouteau, \& Pratt 2005; Vikhlinin et al.\
2006; Schmidt \& Allen 2007; galaxy velocity distributions: Diaferio, Geller, \&
Rines 2005; SZ measurements: Atrio-Barandela et al.\ 2008; strong and weak
lensing measurements: Broadhurst et al.\ 2005a, hereafter B05a; Broadhurst et
al.\ 2005b, hereafter B05b; Limousin et al.\ 2007; Medezinski et al.\ 2007;
Lemze et al.\ 2008, hereafter L08; Broadhurst et al.\ 2008; Zitrin et al.\ 2009,
2010, 2011; Umetsu et al.\ 2010). If both the density and velocity anisotropy
profiles are indeed universal, they must be correlated. Hansen \& Moore (2006,
hereafter HM06), who used N-body simulation, have recently argued for a
universal relation between the DM radial density slope $\gamma(r)$ and the
velocity anisotropy $\beta(r)$ for structures in virial equilibrium. Their
deduced relation was claimed to hold for various systems, including disk galaxy
mergers, simulated halos undergoing spherical collapse, and CDM halos both with
and without cooling.

However, while an analysis of 6 high-resolution simulated galactic halos from
the Aquarius project, carried out by Navarro et al.\ (2010), exhibited a
reasonably good fit to the HM06 relation in the inner regions, large deviations
were reported outside $r_{-2}$, the radius at which the profile slope reaches
$-2$. Analogous results were obtained in a study conducted by Tissera et al.\
(2010), in which they resimulated 6 (Aquarius) galactic halos, 
constructed so as to include metal-dependent cooling, star formation, and 
supernova feedback. In 3 of the halos a rather good match to the HM06
relation was found at small radii, $2\,\,kpc\cdot h^{-1}<r<r_{-2}$, but 
no corresponding match was found in the other 3 halos. No evidence is seen for
the HM06 relation at large radii, $r>r_{-2}$, in any of the six halos. Ludlow
et al.\ 2011, who used cosmological N-body simulations, analyzed relaxed halos
with mass above $\sim 3.4 \times 10^{12}$ h$^{-1}$ M$_{\odot}$ and fitted an
Einasto (1965) profile to the mass density profile. They found a good agreement
to the HM06 relation in the inner regions, i.e. $r<r_{-2}$, and for profiles
with high Einasto logarithmic slope values, $\gtrsim 0.2$, a better
agreement with the HM06 relation achieved even for larger radii. Such a relation
between the DM density and velocity anisotropy is of interest, since if it
exists in relaxed systems, as claimed in previous works (e.g. Hansen \& Stadel
2006; Hansen 2009), this can be an indicator for the system relaxation level. In
addition, since DM velocity anisotropy is not easily measurable, whereas the
density profile can be determined in several different ways based on different
sets of measurements, we can use the $\gamma(r)$ - $\beta(r)$ relation to infer
the DM velocity anisotropy from the density profile.

We report the results of an analysis of a large number of cluster-size halos 
drawn from an Adaptive Mesh Refinement (AMR) cosmological simulation (for
details see \textsection~\ref{The simulation}). The large
number of halos at different redshifts allows us to address the dependence of
the DM velocity anisotropy profile on redshift, halo mass, degree of relaxation,
modeled both as spherical and triaxial DM configurations, and to address also 
the $\gamma$-$\beta$ relation. The outline of the paper is as
follows. In \textsection~\ref{The simulation} we describe the simulation
dataset, and in \textsection~\ref{Radial profiles} we describe how we infer the
radial profiles of the density and velocity anisotropy in spherical and
elliptical shells. In \textsection~\ref{Criteria for relaxed clusters} we
specify our criteria for relaxed halos, and in \textsection~\ref{Results} we
present halos ellipticities at different relaxation levels, the $\beta$
profiles for different halo mass, redshift, and relaxation stages, and the
deduced $\gamma - \beta$ relation. Then in \textsection~\ref{Discussion} we
discuss our findings, and we conclude in \textsection~\ref{Conclusions}.

\section{The simulation} 
\label{The simulation}
 
Clusters of galaxies were found using the HOP halo-finding algorithm (Eisenstein
\& Hut 1998) from a cosmological AMR simulation performed
with the hydrodynamical ENZO code developed by Bryan \& Norman (1997; see also
Norman \& Bryan 1999; Norman et al.\ 2007), assuming a spatially flat
$\Lambda$CDM model with the parameters $\Omega_m = 0.3$, $\Omega_b = 0.04$,
$\Omega_{CDM} = 0.26$, $\Omega_\Lambda = 0.7$, $h=0.7$ (in units of 100
km/s/Mpc), and $\sigma_8 = 0.9$. We also used an Eisenstein \& Hu (1999) power
spectrum with a spectral index of $n = 1$. The hydrodynamics in the AMR
simulation used an ideal gas equation of state (i.e., neither radiative heating,
cooling, star formation or feedback were included), with a box size of 512
h$^{-1}$ Mpc comoving on a side with $512^3$ DM particles, and DM mass
resolution of about $10^{11}$ h$_{0.7}^{-1}$ M$_{\odot}$. The root grid
contained $512^3$ grid cells, and the grid was refined by a factor of two, up to
seven levels, providing a maximum possible spatial resolution of $7.8$
$(1+z)^{-1}$ h$^{-1}$ kpc (this resolution is dependent on the criteria for
refinement of the adaptive mesh, and we used the actual resolution when
analyzing the halos). For more details on the simulation setup and analysis, see
Hallman et al.\ (2007), in particular Section 2.2. 

To find the desired halo DM properties we extracted particle positions and 
velocities from the raw data. Particles within a cube with comoving side of 16
h$^{-1}$ Mpc were extracted. This ensured that both the halo and a sufficiently
large surrounding region was available for examination. We had $6019$, $1391$,
and $69$ halos with $M_{\rm vir}\geqslant10^{14}$ h$_{0.7}^{-1}$ M$_{\odot}$ at
$z=0$, $z=1$, and $z=2$, respectively. 

\section{Radial profiles}
\label{Radial profiles}

\subsection{Measuring halos shape}
\label{Measuring halos shape}
Radial profiles were extracted in both spherical and triaxial shells. The
mass distribution is described in terms of the axial ratios of the density 
surface contours. Assuming that the density distribution is stratified in
similar ellipsoids, it is possible to determine the axial ratios without
knowledge of the radial density distribution (Dubinski \& Carlberg 1991, but
see also other works e.g. Katz\ 1991; Warren et al.\ 1992; Jing et al.\ 1995;
Jing \& Suto\ 2002; Allgood et al.\ 2006, and references therein). 
The mass density in a triaxial configuration, $\rho \equiv \rho(r_{\rm e})$, is
specified in terms of the elliptical distance in the eigenvector coordinate
system of the halo particles, $r_{\rm e}$, 
\begin{equation}
r_{\rm e} = \left(x^2+\frac{y^2}{q^2}+\frac{z^2}{s^2} \right)^{1/2} \,
\label{r_e}
\end{equation} 
where $q$ and $s$ are the normalized axial ratios with $s \leqslant q 
\leqslant 1$. 

These ratios can be derived from the tensor
\begin{equation}
M_{\rm ij} = \sum \frac{x_{\rm i}x_{\rm j}}{r_{\rm e}^2} \,
\end{equation} through 
\begin{equation}
q = \left( \frac{M_{\rm yy}}{M_{\rm xx}} \right)^{1/2} \;\;\; {\rm and}
\;\;\;s = \left( \frac{M_{\rm zz}}{M_{\rm xx}} \right)^{1/2} \,
\end{equation} where the sum is over all the particles, and  
$M_{\rm xx}$, $M_{\rm yy}$, and $M_{\rm zz}$ are the principal components of 
the diagonalized tensor, with $M_{\rm zz}\leqslant M_{\rm  yy}\leqslant 
M_{\rm xx}$. An advantage of this scheme is the equal weighting given to each
particle irrespective of its radial position (see Zemp et al.\ 2011). The large
number of particles in each halo allows accurate determination of the axial
ratios. In practice, the value of $r_{\rm e}$ in $M_{\rm ij}$ is not known in
advance, due to its dependence on $q$ and $s$ through eq.~\ref{r_e}.
The axial ratios are therefore determined iteratively. $M_{\rm ij}$ is initially
calculated assuming that the contours are spherical, so that $q=s=1$.
Particle positions are first rotated into the diagonalized frame of $M_{\rm
ij}$, where only particles inside of the ellipse volume were taken (a sphere,
in the first iteration). The values of $q$ and $s$ are determined from $M_{\rm
ij}$ and then used to recalculate $r_{\rm e}$ in this new frame and fed back
into the $M_{\rm ij}$ relation to determine iterated values of $q$ and $s$. When
the input values match the output values within a certain tolerance, convergence
to the true axial ratios is achieved.

In each iteration new values for $q$ and $s$ are determined, so the halo volume
is deformed. We kept the magnitude of the semi-major axis equal to $r_{\rm
vir}$ of the original spherical radius, i.e. $r_{\rm e} = r_{\rm vir,
spheric}$. This radius was chosen and not a longer one, so the ellipticity will
not be affected by closeby clumps. Thus, during the volume deformation only
the two smaller axes were changed. Then we took halos which their ellipsoid
volume contained a large number of particles, $> 10^{3}$ (which gave $3069$,
$364$, and $13$ halos at $z=0$, $z=1$, and $z=2$, respectively). 

\subsection{Radial velocity profiles}

The DM velocity anisotropy profile for each halo was determined as follows: we
first identified the halo center with the peak of the surrounding 3D density
distribution and then determined the proper (non-comoving) velocities of the
DM particles with respect to the cluster center by subtracting the velocity of
the halo center. This procedure was carried out for 15 equally spaced shells
within the virial radius, a division that yields DM particle counts of the same
order of magnitude in each bin. Logarithmic spacing was impractical due to the
low spatial resolution. The DM velocity anisotropy in each shell was calculated
as 
\begin{equation}
\beta = 1-\frac{\sigma_{\theta}^2+\sigma_{\phi}^2}{2\sigma_r^2}\ ,    
\end{equation} 
where $\sigma_r$, $\sigma_{\theta}$, and $\sigma_{\phi}$ denote the radial,
polar, and azimuthal velocity dispersions, respectively. The velocity
dispersion is defined as follows $\sigma_{\rm i} \equiv
\sqrt{<v_i^2>-<v_i>^2}$, where $i=\theta$, $\phi$, and $r$. Shells
containing less than 10 DM particles were excluded by virtue of their
statistical insignificance. We only considered halos containing at least $10^3$
particles, so as to obtain robust results independent of numerical artifacts (as
has also been done by Neto et al.\ 2007). Since our DM mass resolution is
approximately $10^{11}$h$_{0.7}^{-1}$ M$_{\odot}$, we examined all halos having
$M_{\rm vir}\geqslant10^{14}$ h$_{0.7}^{-1}$ M$_{\odot}$. We also only
considered halos with $q>0.4$, since $q<0.4$ values are due to closeby
structures, (which gave $2969$, $348$, and $13$ halos at $z=0$, $z=1$, and
$z=2$, respectively).

\subsection{Radial mass density profiles}
For constructing the DM density profile we used the same binning and 
halo center definition as in DM velocity anisotropy profile, and
averaged the DM density over spherical shells. The radial density 
slope is defined as 
\begin{equation}
\gamma(r)=\frac{d\ln[\rho_{\rm DM}(r)]}{d\ln[r]}\ .
\end{equation} 
For comparison with the radial density slope derived from a fit to 
an NFW profile, we fitted the resulting distribution to an NFW profile, 
$\rho_{i}^{NFW} = \frac{4\rho_s} {\left( r_i/r_s\right )\left( 1+r_i/r_s 
\right)^2}$, where $r_s$ and $\rho_s$ are a scale radius and the density 
at this radius, respectively, both of which were treated as free parameters.
For estimating the virial radius (both in the spherical and elliptical
cases), the value for the final overdensity to the critical density at collapse
was taken to be $\Delta_c = 18\pi^2+82x-39x^2$, where $x \equiv \Omega_m(z)-1$
when $\Omega_m(z)$ is the ratio of the matter density to the critical
density (Bryan \& Norman 1998). Note, the elliptical virial radius is obviously
larger than the spherical virial radius. The best fit was found by
minimizing
\begin{equation}
\chi^2=\sum_{i=1}^{N_{bins}} \left[
\frac{\log(\rho_i)-\log(\rho_{i}^{NFW}(\rho_0,r_s))}{\log(
\sqrt{1/N_i}\rho_i)} \right]^2 \ , 
\end{equation} where each bin was assigned weight by the bin particle number,
$\Delta \rho_i/\rho_i  = \sqrt{N_i}$, so bins with fewer particles get lower
weight.

The virial radius was estimated to be at the radius where $\overline{\rho} =
\Delta_c \rho_{\rm crit}$, where $\rho_{\rm crit}$ is the critical density. The
value was consistent in less than 1\% on average with the value estimated using
the NFW best-fit parameters.

\section{Criteria for relaxed clusters}
\label{Criteria for relaxed clusters}

The distinction between relaxed and unrelaxed clusters was made according to
five criteria some of which were laid down by Thomas et al.\ (2001) and Neto et
al.\ (2007). These are based on the following quantities:
\begin{enumerate}
  
\item The displacement between the center of mass, $r_{\rm cm}$, and the
potential minimum, $r_{\rm p}$, with the latter quantity calculated including 
particles within the virial radius. The displacement was normalized with
respect to the spherical virial radius, $r_{\rm offset}=|r_{\rm p}-r_{\rm
cm}|/r_{\rm vir}$ (see also Neto et al.\ 2007).
  
\item The virial ratio, $2T/|U|$ (see also Neto et al.\ 2007). We computed the
total kinetic and gravitational energies of the halo particles within
$r_{\rm vir}$.
When halos were modeled as triaxial, their major axes were set equal to the
virial radii of the respective spherical configurations. For the estimation of
$T$ we subtracted the motion of the halo center, whereas $U$ was calculated
using a random sample of 1000 particles. We estimated the precision level of
this method by (a) repeating the calculation 10 times for the most massive halo
(the one containing the largest number of DM particles), which generated a
relative difference of $(1.4 \pm 1)\%$, and (b) calculating $U$ in a single
halo, using $10^4$ particles. The relative average difference produced by this
method was $0.8\%$.

\item The corrected virial ratio, $(2T-Es)/|U|$. For details about this
criterion see \textsection~\ref{Correction for the virial ratio}.

\item The displacement between the density peak $r_{\rm d}$ and the center of
mass $r_{\rm cm}$, with the latter quantity calculated using particles within
the virial radius. The displacement was normalized with respect to the spherical
virial radius, $r_{\rm sub}=|r_{\rm d}-r_{\rm cm}|/r_{\rm vir}$. Thomas et
al.\ (2001) interpreted this displacement as a measure of the substructure
level.

\item The displacement between the density peak $r_{\rm d}$ and the potential
minimum, $r_{\rm p}$, with the latter quantity calculated using particles within
the virial radius. The displacement was normalized with respect to the spherical
virial radius, $r_{\rm dp}=|r_{\rm d}-r_{\rm p}|/r_{\rm vir}$.

\end{enumerate}

In equilibrium $r_{\rm offset}$ would be expected to vanish, the virial
ratio would approach a value slightly higher than unity, since even in relaxed
systems there always is some infalling matter, and the corrected virial ratio
(for more details see \textsection~\ref{Correction for the virial ratio})
should approach unity. While all five criteria are related to the degree of
relaxation in a straightforward manner, the boundary levels between the two
phases are quite arbitrary. For example, for the first two Neto et al.\ (2007)
adopted $r_{\rm offset}=0.07$ and $2T/|U|=1.35$.

In this paper we focused on the first three criteria; however, the correlations
between all of the five criteria and other parameters are shown in
\textsection~\ref{appendix}.

\subsection{Correction for the virial ratio}
\label{Correction for the virial ratio}

Halos are not isolated systems, since matter is continuously falling onto them.
Thus, we also calculated the virial ratio taking under consideration the
infalling matter onto the halos, which was claimed to have a significant overall
contribution to the pressure at the halo boundaries (Shaw et al.\ 2006; Davis,
D'Aloisio, \& Natarajan 2011). This modification manifests itself as a form of
surface pressure at the boundaries of the halos (Chandrasekhar 1961; Voit 2005,
and references within), so in steady state the virial equation is $2T+U-E_{\rm
s}=0$. Here we estimated the surface pressure term in an Eulerian virial theorem
version, since each of the halos has a known (fixed) volume (see McKee \&
Zweibel 1992; Ballesteros-Paredes 2006, and references therein). Thus, its
energy content is
\begin{equation}
E_{\rm s} = \oint \rho \vec{r}\cdot\vec{v} \vec{v}\cdot\vec{dS} = m_{\rm
DM} \sum_{\rm i} \vec{r_{\rm i}} \cdot \vec{v_{\rm i}} v_{\rm n,i} \frac{S}{V}
\,
\end{equation} where the loop denotes integration over a closed surface, and the
summation is over all particles in a shell with surface $S$ and volume $V$. The
symbols $m_{\rm DM}$, $\vec{r_i}$, $\vec{v_i}$, and $v_{\rm n,i}$ are the DM
particle mass, vector position and velocity, and the outward normal component of
$v$, respectively. In practice, we followed Shaw et al.\ and computed $E_{\rm
s}$ in the outer 20\% of the spherical virial (elliptical) radius in the
spherical (elliptical) shells case (repeating the analysis for 10\% of the
virial radius). The innermost, median, and outer shells were taken to be
$0.8R_{\rm vir}$, $0.9R_{\rm vir}$, and $R_{\rm vir}$, respectively.

In the case of spherical symmetry $E_s = 4\pi R^3P_{\rm s}(R)$, since the
integration is carried out over the angles, and the radius is constant. We 
can estimate the surface pressure term as $P_{\rm s} = \frac{m_{\rm
DM} \sum_{\rm i} v_{\rm r,i}^2}{V}$, where $V = \frac{4\pi}{3}[R_{\rm
vir}^3-(0.8R_{\rm vir})^3]$ is the outer 20\% shell volume. Thus, $E_{\rm s} = 3
\frac{0.9^3}{1-0.8^3} m_{\rm DM} \sum_{\rm i} v_{\rm r,i}^2$.

In the case where the halo mass distribution is modeled as an ellipsoid with
isodensity surfaces $\vec{n}= \nabla r_{\rm e}$, where $\vec{n}$ is the
"direction" of the surface ($\vec{dS} \equiv \vec{n}dS$) and $r_e$ is the
expression given in eq.~\ref{r_e}, so $\vec{n} = \nabla r_{\rm e} = r_{\rm
e}^{-1} (x,y/q^2,z/s^2)$, where $x$, $y$, and $z$ are at the particle position.
Thus, $\widehat{n}∗ \equiv \vec{n}/||n|| = (x,y/q^2,z/s^2) /
\sqrt{x^2+y^2/q^4+z^2/s^4}$, and $\vec{v_n} = \widehat{n}∗ \cdot
(v_x,v_y,v_z)$, where this is done for each particle in the elliptical shell at
the elliptical virial radius.

\begin{figure}
\centering
\epsfig{file=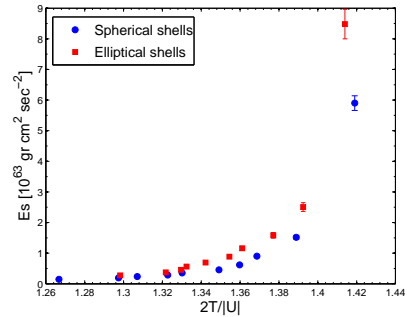, width=6cm, clip=}
\caption{Mean over halo surface pressure term vs. virial ratio at $z=0$. Blue
circles are for spherical shells containing particles within 20\% of the
spherical $r_{\rm vir}$ shell, and red squares are for elliptical
shells containing particles within 20\% of the elliptical $r_{\rm vir}$ shell.
The uncertainty was taken to be Poissonian, $\Delta <E_{\rm s}> = <E_{\rm
s}>/\sqrt{N}$ where $N$ is the number of halos in the bin. 
\label{mean_Es_vs_virial_ratio_spheric_n_elliptic}}
\end{figure}

In figure~\ref{mean_Es_vs_virial_ratio_spheric_n_elliptic} we plotted the
correction term as a function of the virial ratio.
 It is easy to see that the correction is larger in elliptical
shells. Thus, for accurately using this proxy, using spherical shells maybe not
good enough.

\section{Results}
\label{Results}

\subsection{Halos ellipticities at different relaxation levels} 
\label{Halos ellipticities at different relaxation levels}
\begin{figure}
\centering
\epsfig{file=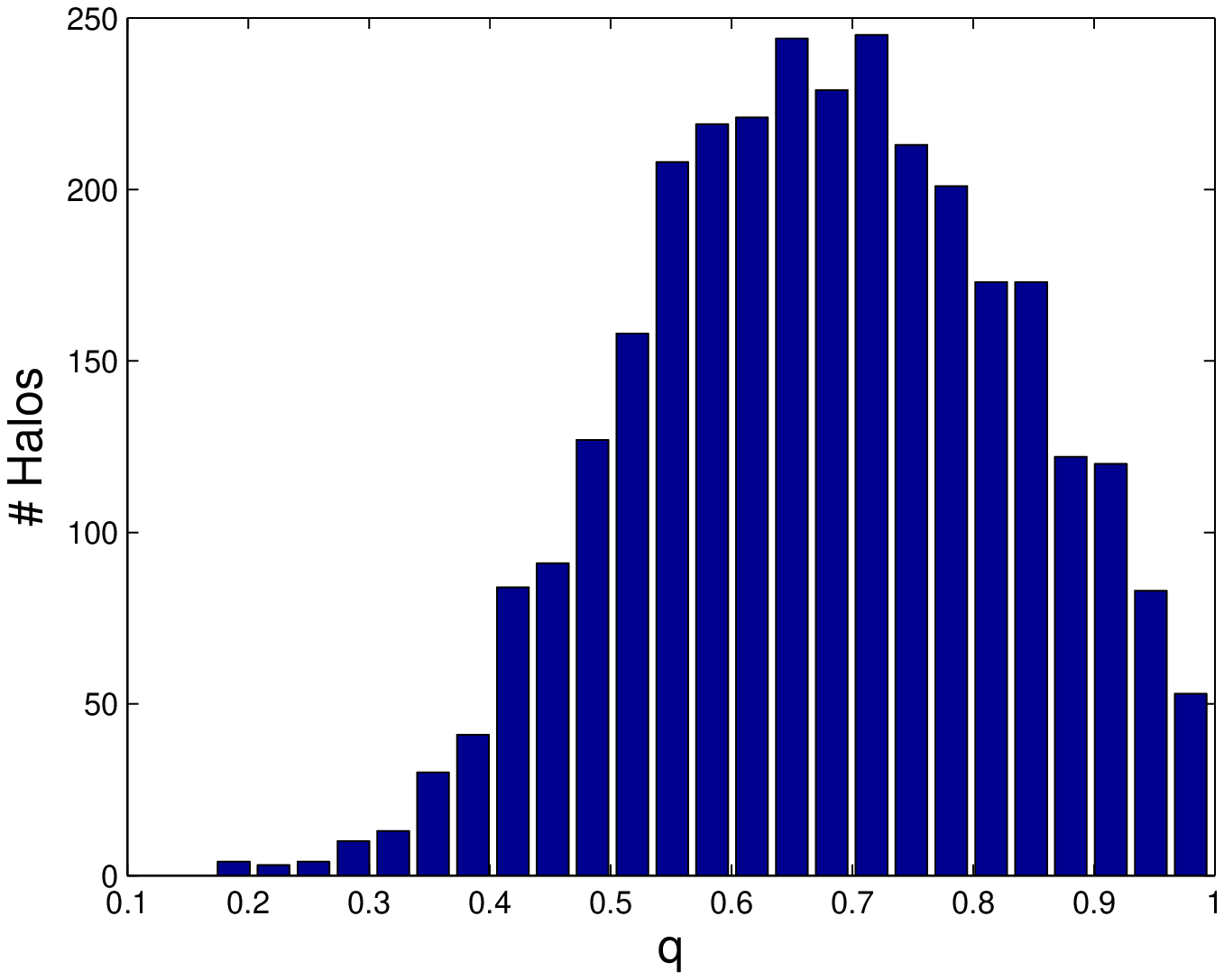, width=6cm, clip=}
\epsfig{file=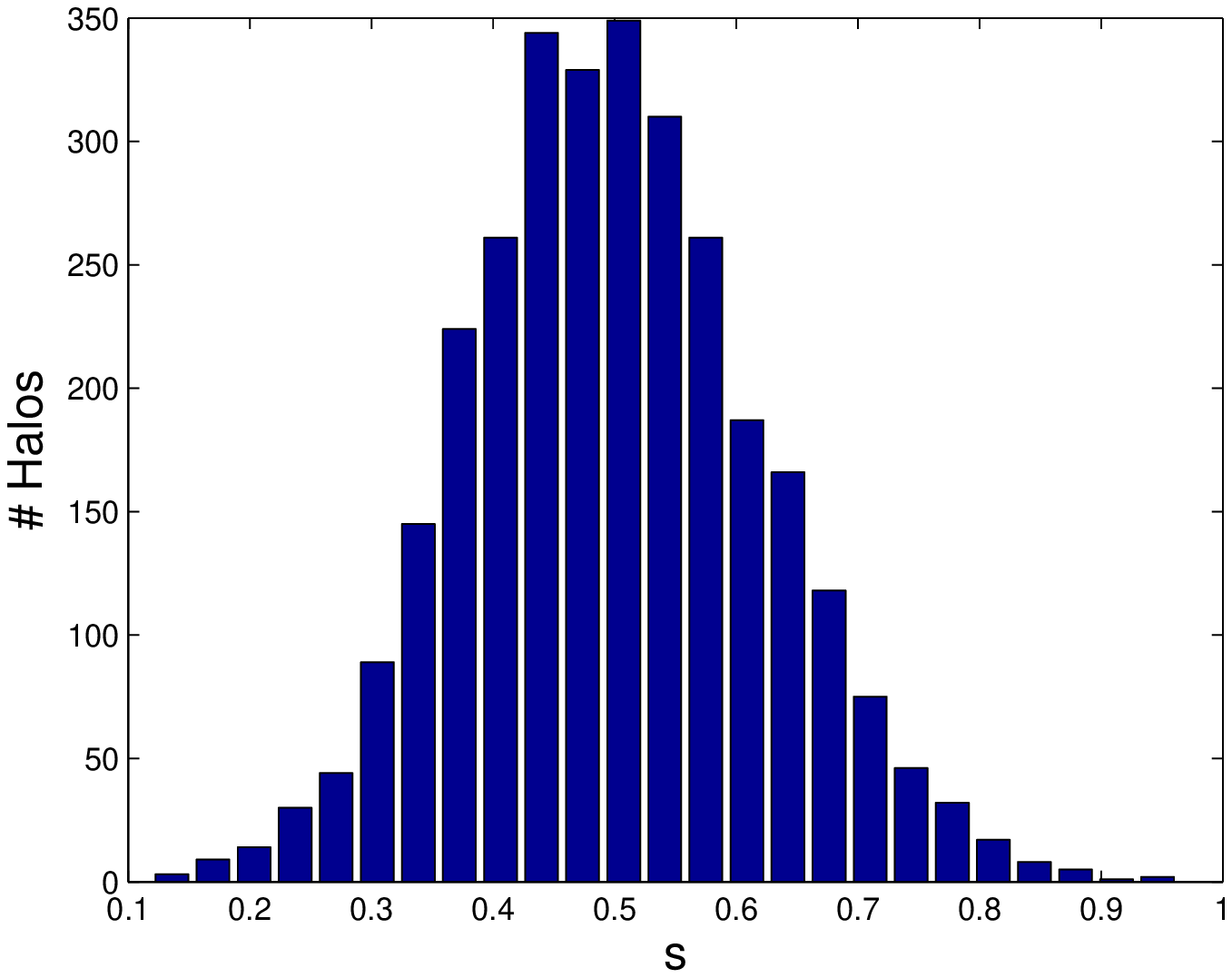, width=6cm, clip=}
\caption{Axes ratio histogram, where top and bottom panels are for $q$ and $s$
histograms, respectively.
\label{q_n_s_histograms}}
\end{figure}

\begin{figure}
\centering
\epsfig{file=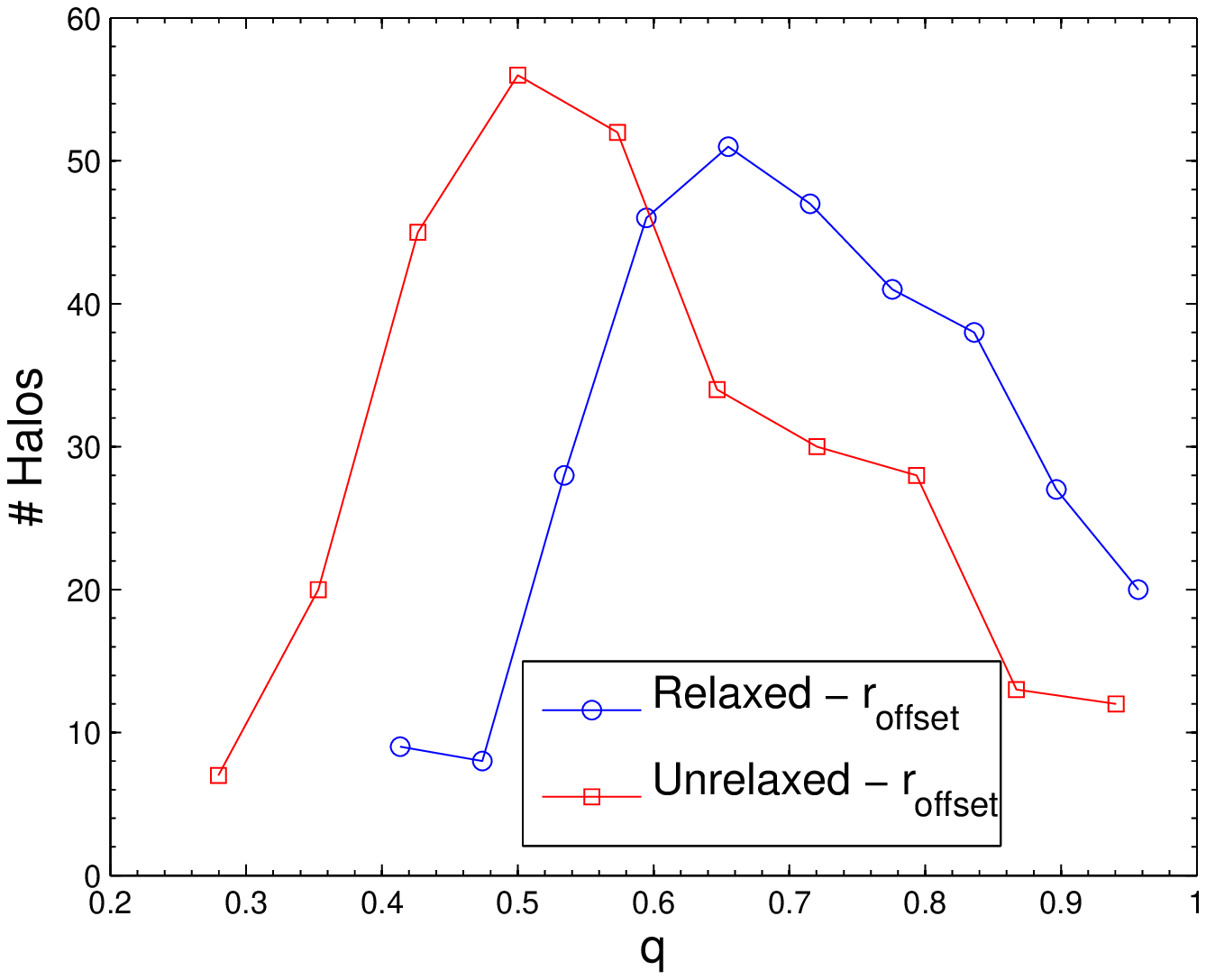, width=6cm, clip=}
\epsfig{file=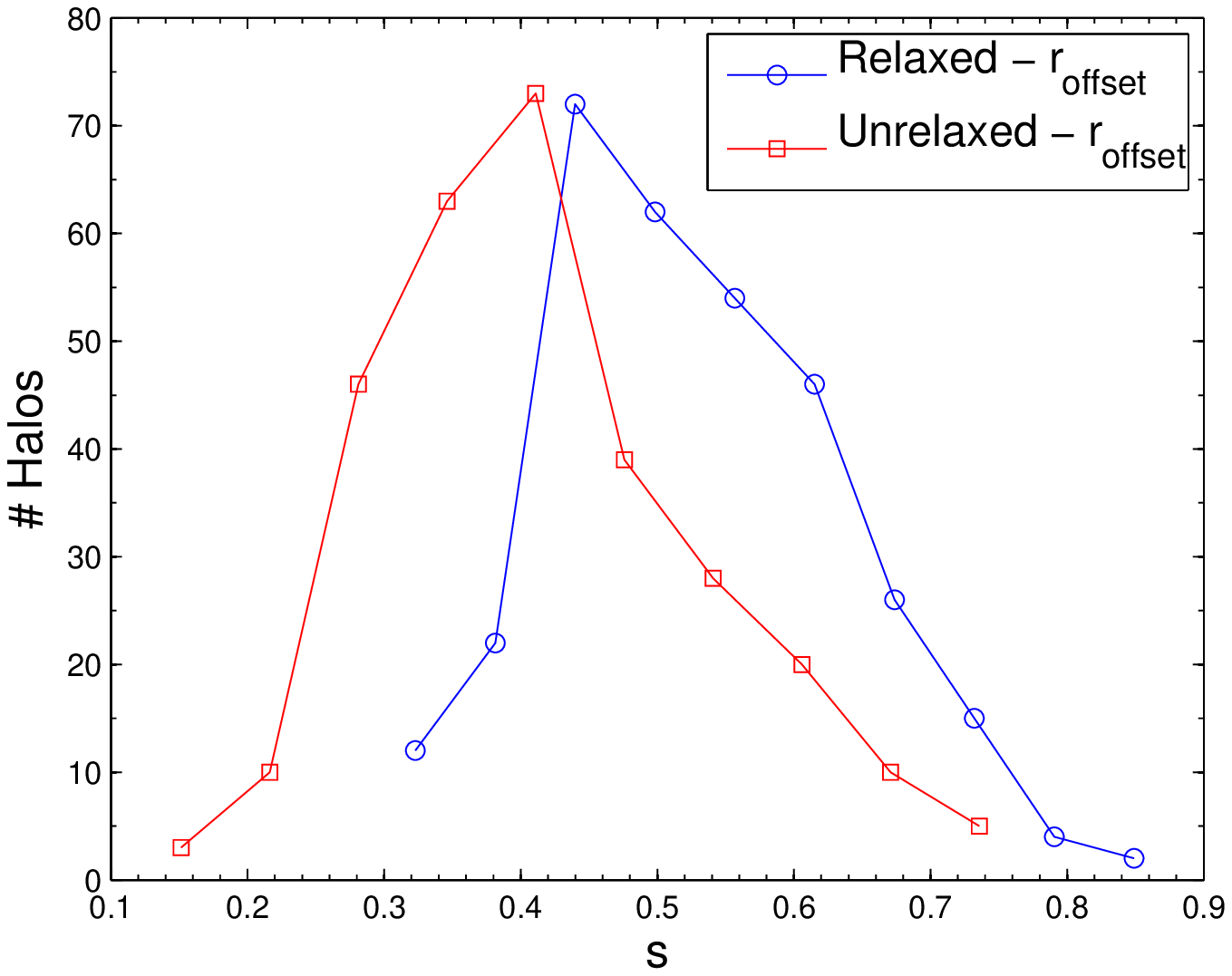, width=6cm, clip=}
\caption{Axes ratio histogram for halos in different relaxation level, when the
distinction is made according to the $r_{\rm offset}$ criterion. The relaxed and
unrelaxed samples have $r_{\rm offset}<0.02$ (315 halos) and $0.12\lesssim
r_{\rm offset}<0.2$ (297 halos), respectively. Upper and lower
panels are for $q$ and $s$ histograms, respectively.
\label{q_n_s_histograms_relaxation_by_s}}
\end{figure}

\begin{figure}
\centering
\epsfig{file=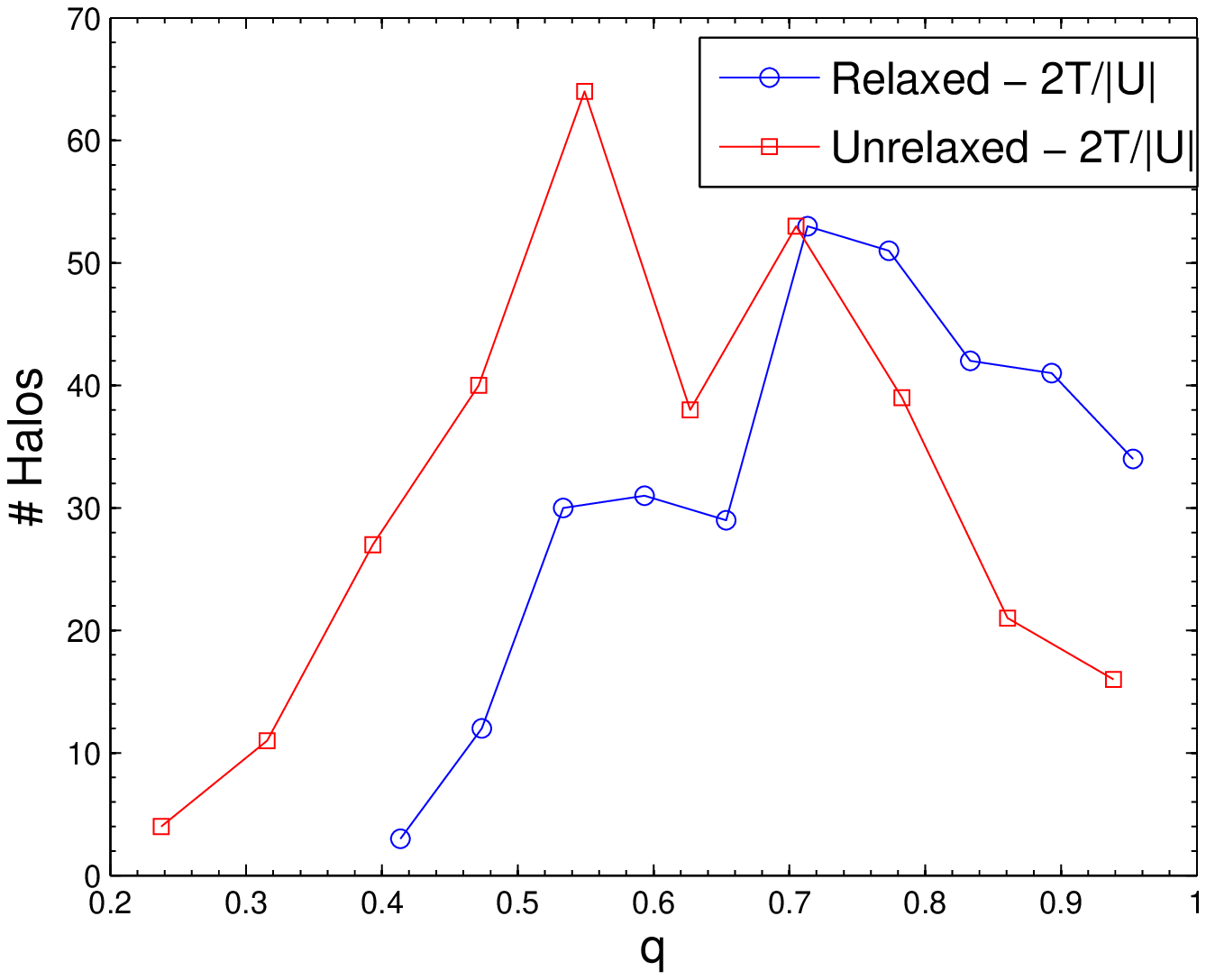, width=6cm, clip=}
\epsfig{file=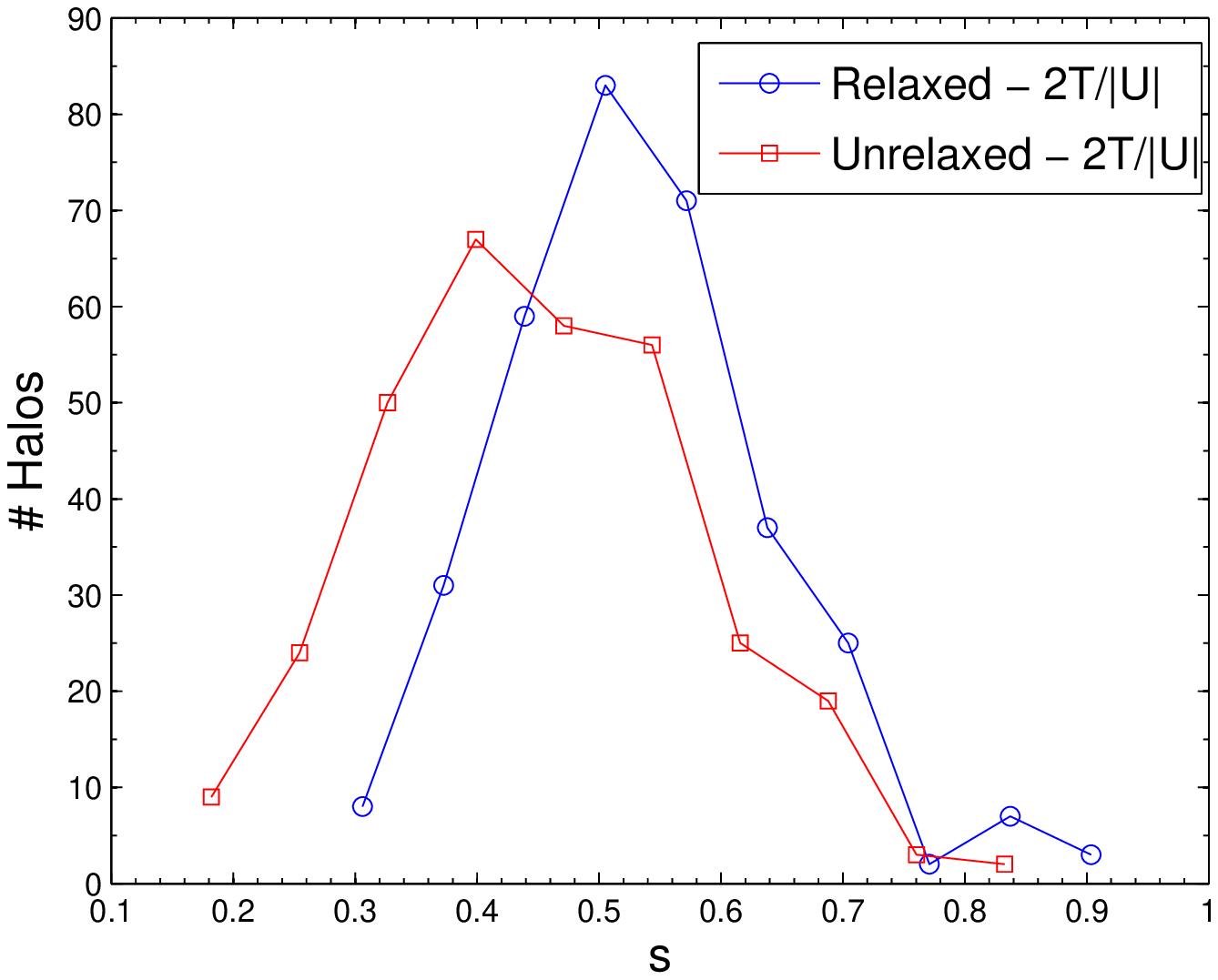, width=6cm, clip=}
\caption{Axes ratio histogram for halos in different relaxation level, when the
distinction is made according to the $2T/|U|$ criterion. The relaxed and
unrelaxed samples have $2T/|U|<1.25$ (326 halos) and $1.44\lesssim 2T/|U|<1.5$
(313 halos), respectively. Upper and lower panels are for $q$ and $s$
histograms, respectively.
\label{q_n_s_histograms_relaxation_by_virial_ratio}}
\end{figure}

\begin{figure}
\centering
\epsfig{file=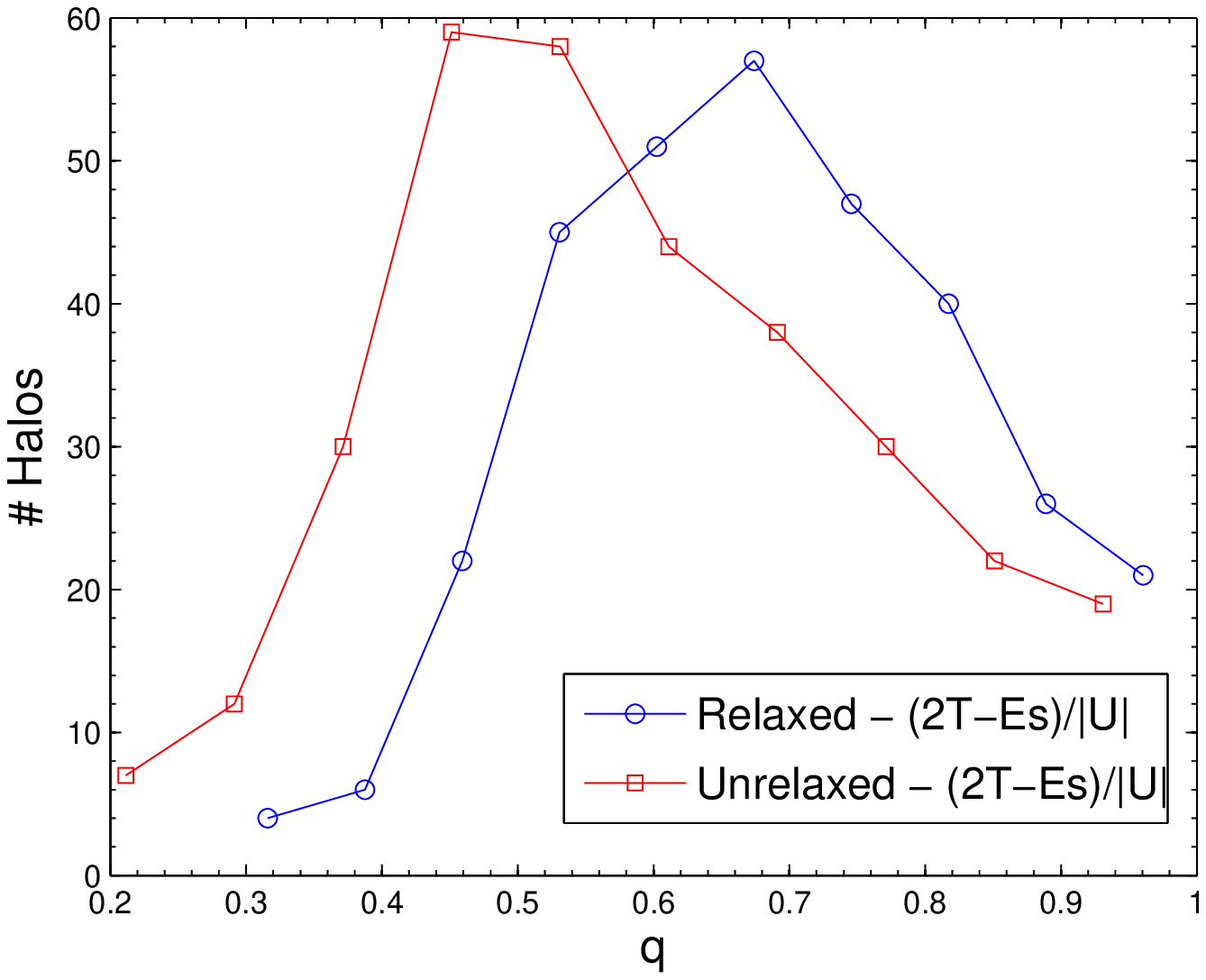, width=6cm,
clip=}
\epsfig{file=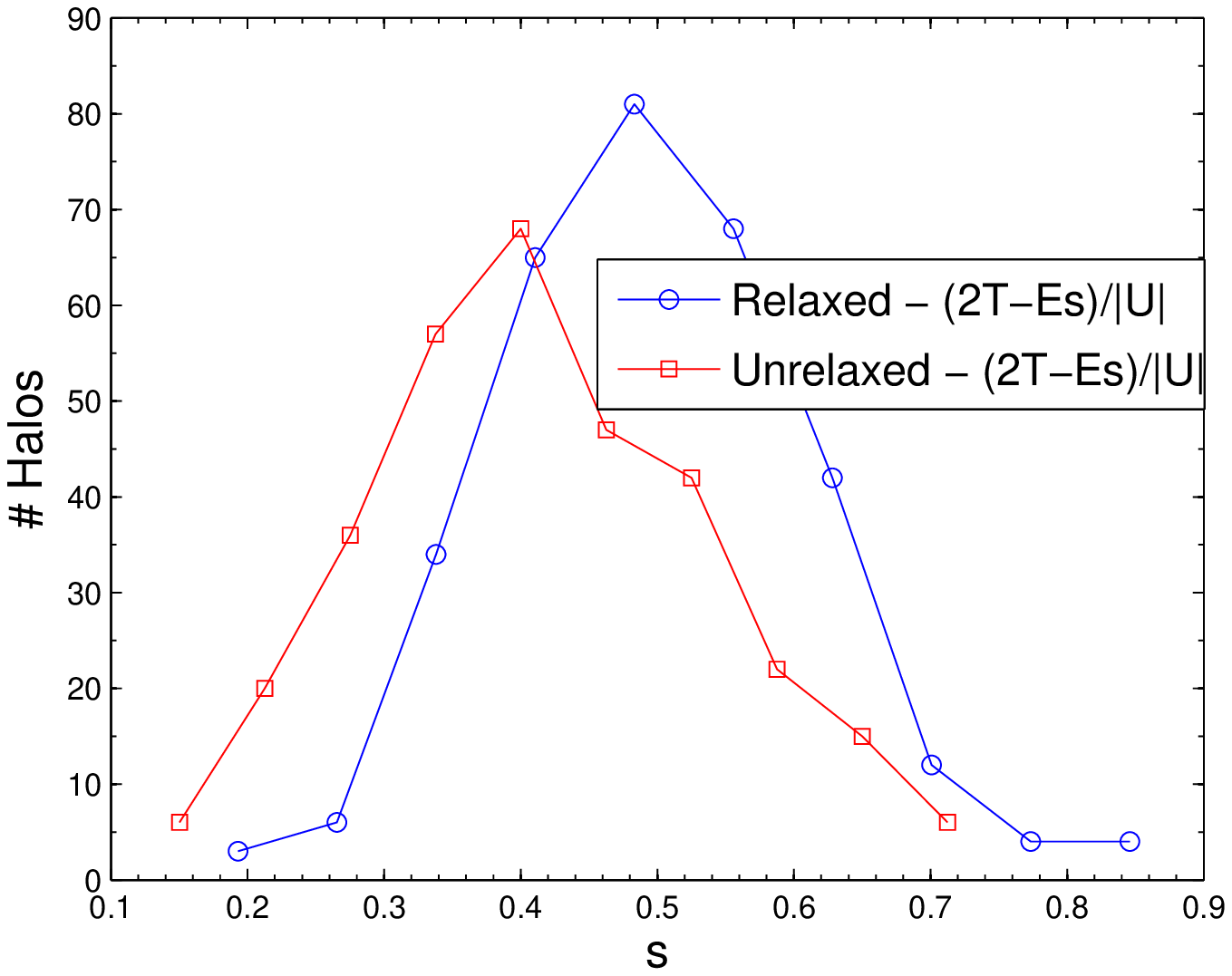, width=6cm,
clip=}
\caption{Axes ratio histogram for halos in different relaxation level, when the
distinction is made according to the $(2T-Es)/|U|$ criterion. The
relaxed and unrelaxed samples have $(2T-Es)/|U|<0.95$ (319 halos) and
$1.22\lesssim (2T-Es)/|U|<1.5$ (319 halos), respectively. Upper and lower
panels are for $q$ and $s$ histograms, respectively.
\label{q_n_s_histograms_relaxation_by_corrected_virial_ratio}}
\end{figure}

The axes ratios, $q$ and $s$, histograms for all halos are plotted in
figure~\ref{q_n_s_histograms}. In
figures~\ref{q_n_s_histograms_relaxation_by_s},
\ref{q_n_s_histograms_relaxation_by_virial_ratio}, and
\ref{q_n_s_histograms_relaxation_by_corrected_virial_ratio} the relaxed and 
unrelaxed halos ellipticities are plotted for the three relaxation criteria,
$r_{\rm offset}$, $2T/|U|$, and $(2T-Es)/|U|$, respectively. The values of the
relaxation criteria were chosen such that in each plot both relaxed and
unrelaxed samples have about the same number of halos, so the histograms will
have a similar normalization. 

\begin{figure}
\centering
\epsfig{file=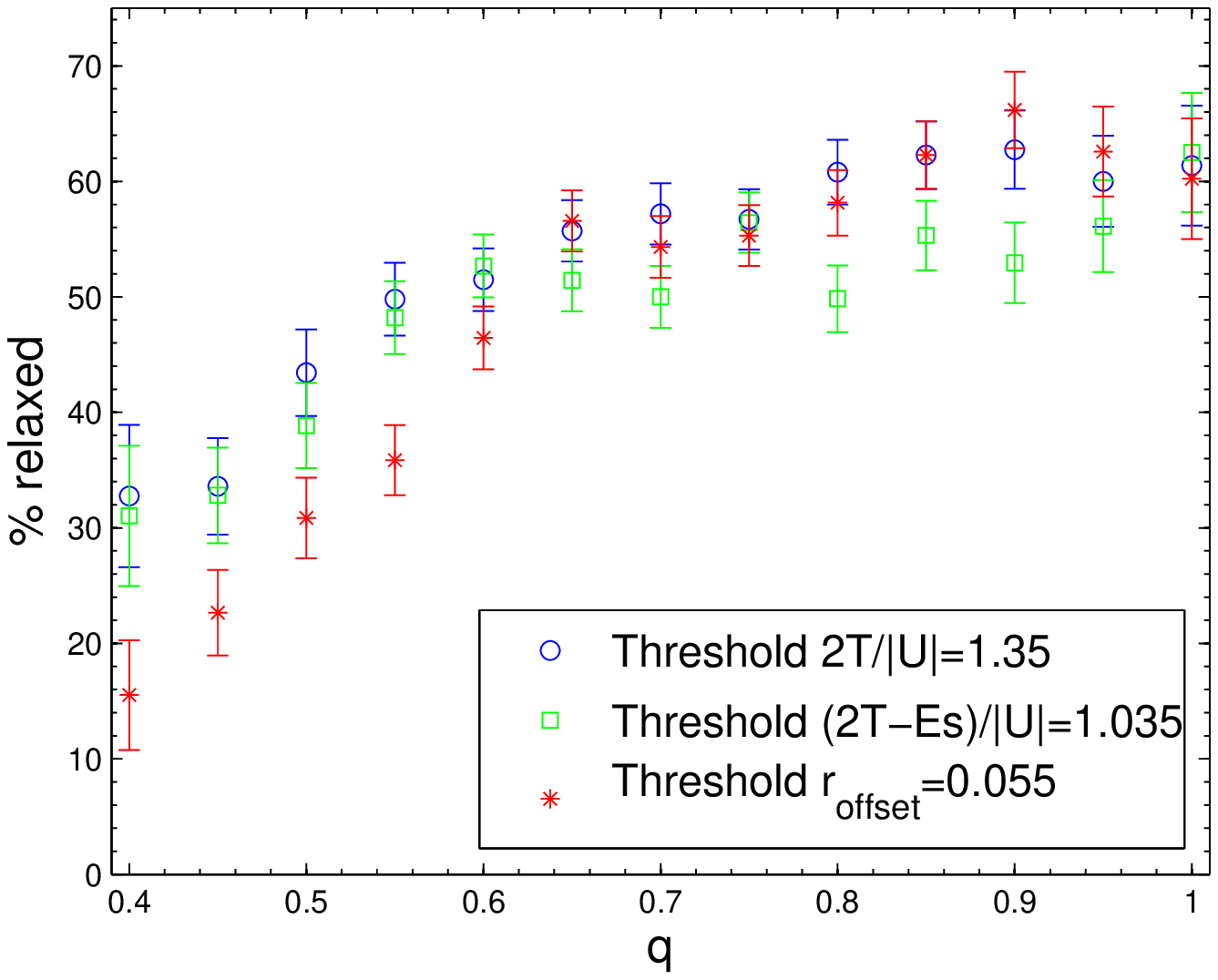, width=8cm, clip=}
\caption{Fraction of relaxed halos at different ellipticities and for
different relaxation criteria, $2T/|U|=1.35$ (blue circles), $(2T-Es)/|U|=1.035$
(green squares), and $r_{\rm offset}=0.055$ (red asterisks). For these
relaxation
estimations we took halo particles within the elliptic virial radius. The
uncertainties were taken to be Poissonian, 
i.e. $\Delta N_{\rm relaxed}= \sqrt{N_{\rm relaxed}}$ and $\Delta N_{\rm
unrelaxed} = \sqrt{N_{\rm unrelaxed}}$.
\label{relaxed_unrelaxed_ratio_vs_q}}
\end{figure}

Similar to figures~\ref{q_n_s_histograms_relaxation_by_s} -
\ref{q_n_s_histograms_relaxation_by_corrected_virial_ratio}, in
figure~\ref{relaxed_unrelaxed_ratio_vs_q} we plotted the fraction of relaxed
clusters, $N_{\rm relaxed}/(N_{\rm relaxed}+N_{\rm unrelaxed})$, vs halo
ellipticity. Here we took only halos with $q>0.4$ since more elliptical halos
are rare and therefore give poor statistics. In addition, for halos with $q<0.4$
the values of both relaxation criteria are strongly dependent on the length of
the ellipse major axis, which is likely due to the fact that many of them are in
the process of a major merger and highly unrelaxed. The threshold was chosen so
the three will have about the same normalization at $q = 1$.
 The fraction of relaxed halos at different ellipticities is a
little bit more sensitive to $r_{\rm offset}$ than to $2T/|U|$ and
$(2T-Es)/|U|$.

\begin{figure*}
\centering
\makebox{
\epsfig{file=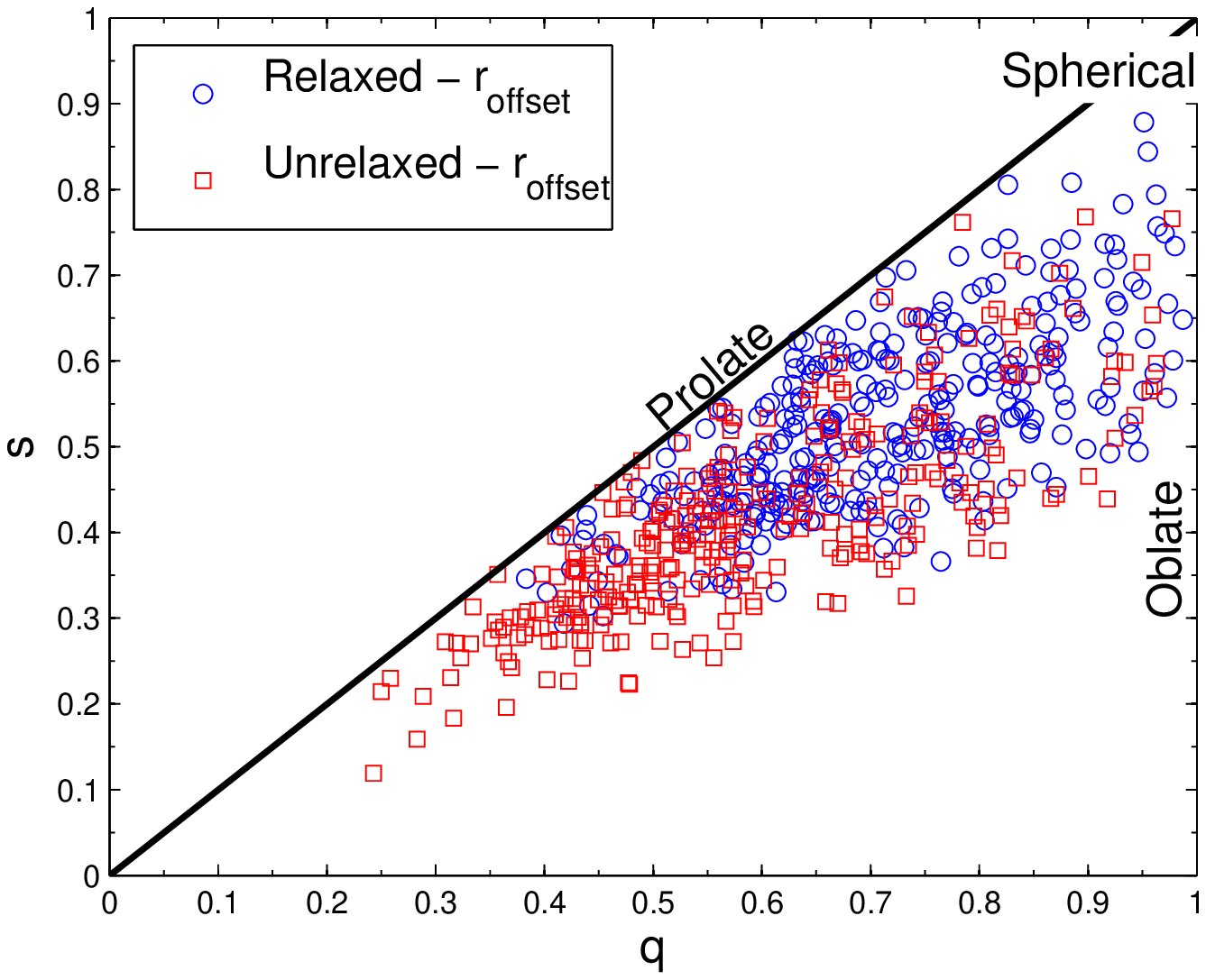, width=6cm,
clip=}
\epsfig{file=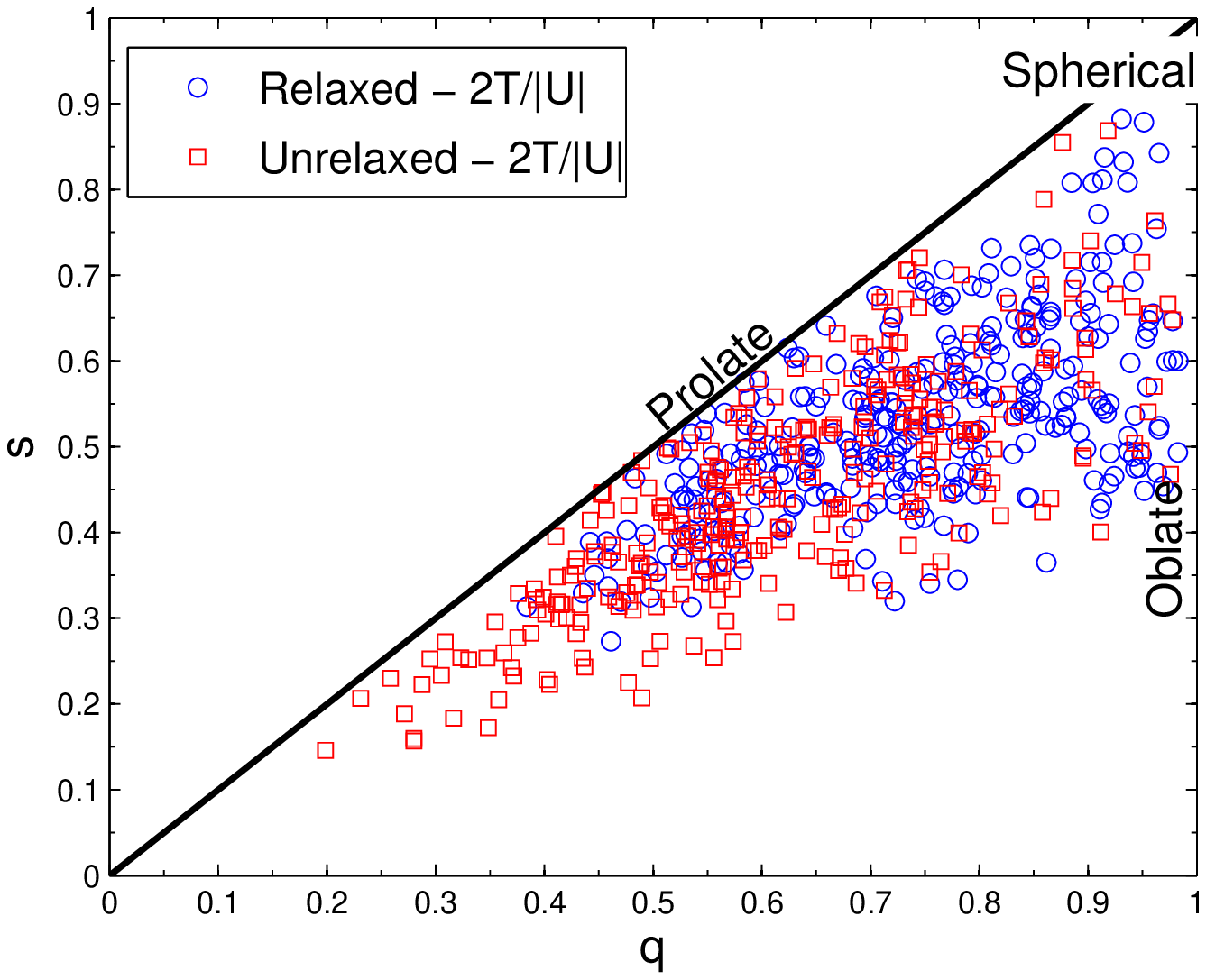, width=6cm, clip=}
\epsfig{file=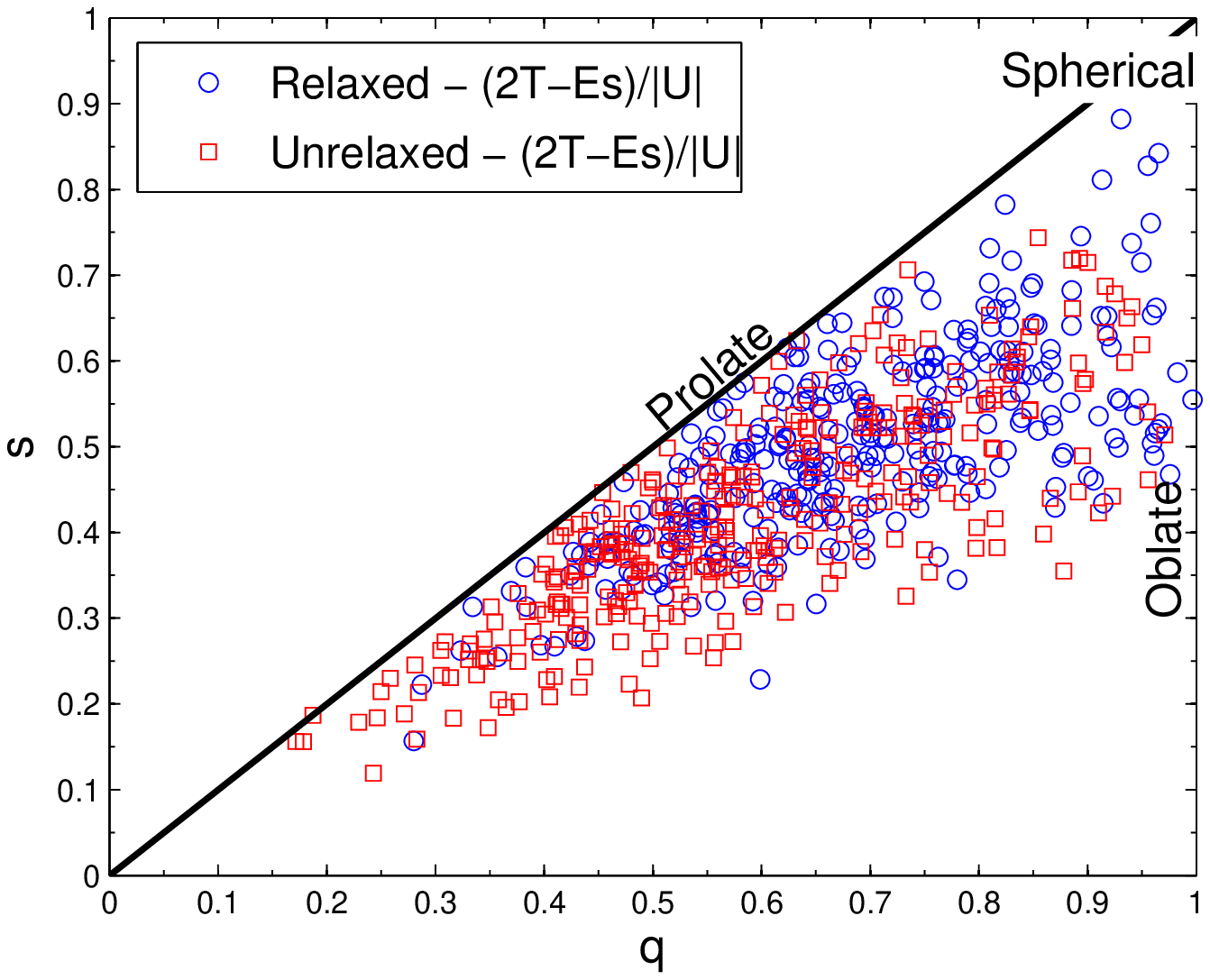, width=6cm,
clip=}
}
\caption{Triaxiality in different relaxation level and for different relaxation
criteria. Left panel: Relaxation gauged by $r_{\rm offset}$, and the threshold
value is the same as figure~\ref{q_n_s_histograms_relaxation_by_s}. Middle
panel:
Relaxation gauged by $2T/|U|$, and the threshold value is the same as
figure~\ref{q_n_s_histograms_relaxation_by_virial_ratio}. Right panel:
Relaxation gauged by $(2T-Es)/|U|$, and the threshold value is the same as
figure~\ref{q_n_s_histograms_relaxation_by_corrected_virial_ratio}. The black
solid line is for $q = s$. All halos are on one side of the black line
since $q>s$ by definition (see \textsection~\ref{Measuring halos shape}).
\label{Triaxiality in different relaxation level}}
\end{figure*}

In figure~\ref{Triaxiality in different relaxation level} we plotted $q$ vs
$s$ at different relaxation level when relaxation is gauged by $r_{\rm
offset}$, $2T/|U|$, and $(2T-Es)/|U|$ for the left, middle, and right panels,
respectively. This is a convenient way to see the differences in the halos 3D
shape between relaxed and unrelaxed halos.

\subsection{Averaged ellipticities vs main axes length}

\begin{figure}
\centering
\epsfig{file=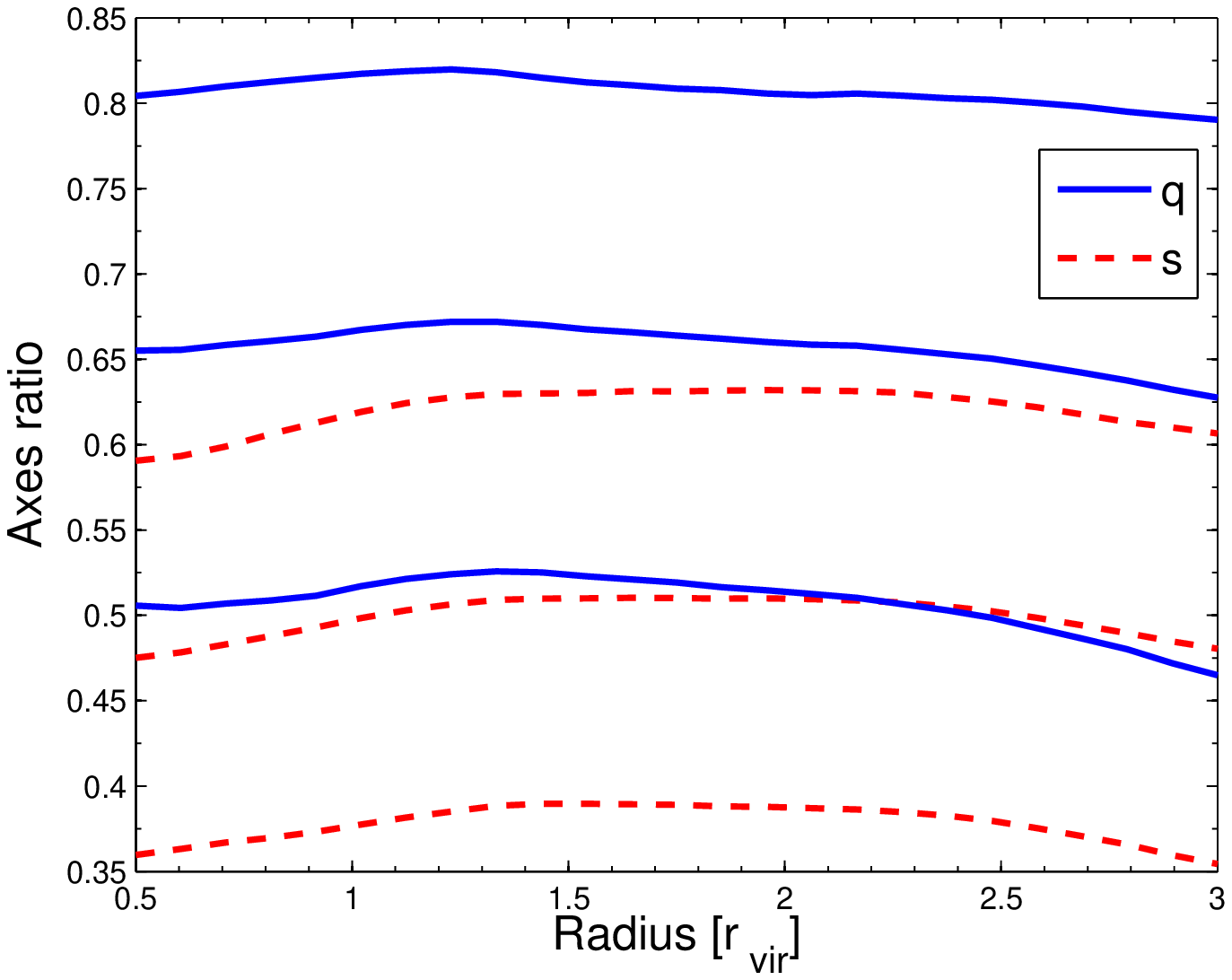, width=8cm, clip=}
\epsfig{file=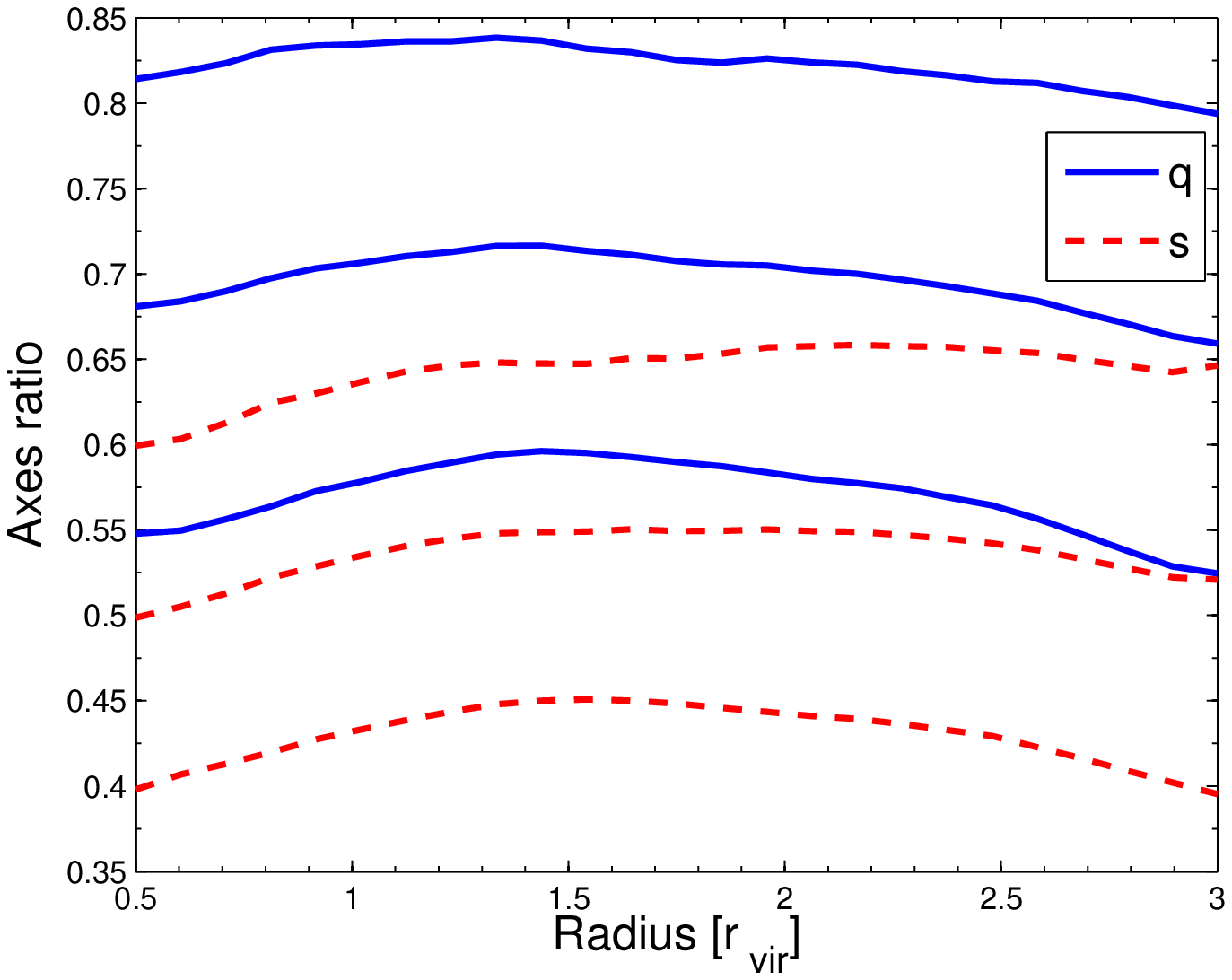, width=8cm, clip=}
\caption{Profiles of the average axes ratio at $z=0$. Top panel: plotted are
the average values of $q$ (solid curve) and $s$ (dashed curve), over all halos
at volumes with a different semi-major axes value at $z=0$. Bottom panel: the
same but only for relaxed halos, $r_{\rm offset} < 0.02$. The central
curves represent the mean value, with the $\pm$ 1--$\sigma$ width of the
distribution uncertainty around the mean is marked by the upper and lower
curves.
\label{q_n_s_diff_radii_z0}}
\end{figure}

In figure~\ref{q_n_s_diff_radii_z0} (top panel) we plotted the averaged $q$ and
$s$ values over halos at $z=0$ at different portions of the virial radius
for the semi-major axis. Only halos with a large number of particles, $>
10^{3}$, inside the smallest radius, $0.5 r_{\rm vir}$, were taken, which gave
$\sim 10^3$ halos. We also plotted (bottom panel) the axes ratios for relaxed
halos, $r_{\rm offset} < 0.02$, which gave us 115 halos. The halo ellipticity
first decreases a small amount, until $\sim (1.5-2)r_{\rm vir} $, then their
ellipticity increases. In other words on average the halos are more elliptical
at small radii then with increasing radius they become more spherical, and then
elliptical again. This behavior is a little bit more pronounce in relaxed
halos. However, the change is small, even in the relaxed sample, and over the
radius range $(0.5-3)r_{\rm vir}$ the axes ratios are quite constant,
$<q>\approx 0.66$ and $<s>\approx 0.5$ and $<q>\approx 0.7$ and $<s>\approx
0.54$, for the whole $\sim 10^3$ halo sample and for the relaxed one,
respectively, especially compared to the large scatter. 

\subsection{The DM velocity anisotropy profile}

\begin{figure}
\centering
\epsfig{file=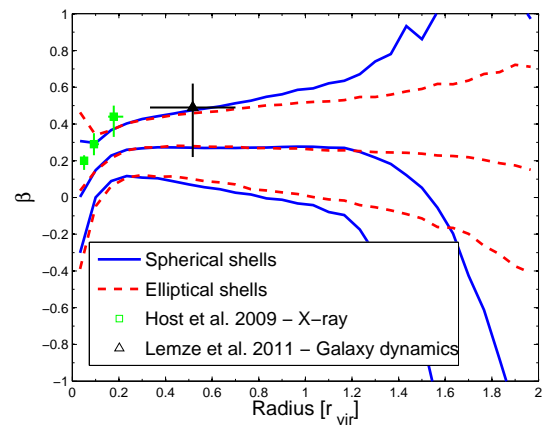, width=8cm,
clip=}
\caption{Velocity anisotropy profiles when the shells are described as
spherical (red dash curves) or elliptical (blue solid curves). Note, the
radius in the elliptical case is the semi-major axis. The central curves
represent the mean value, with the $\pm$ 1--$\sigma$ uncertainty around the
mean is marked by the upper and lower curves. The green squares are averaged
values of 16 clusters, when $\beta$ was inferred from X-ray observations (Host
e.t al.\ 2009). The black triangle is a value inferred for A1689 using galaxy
dynamics information (Lemze et al.\ 2011).
\label{Different shells}}
\end{figure}

To assess the impact of an aspherical halo configuration, we plot in 
figure~\ref{Different shells} the velocity anisotropy profiles in both 
spherical and elliptical shells for all high-mass halos ($M_{\rm
vir}\geqslant 10^{14}$ h$^{-1}_{0.7}$ M$_{\odot}$).  
 It is interesting to note that the average $\beta$ profile
is almost the same in spherical and elliptical shells till $\sim (0.7-1) r_{\rm
vir}$. Beyond this region, however, the scatter of $\beta$ using elliptical
shells is much smaller. On top of the theoretical expectation we plotted the
$\beta$ values estimated indirectly from data. The green squares are averaged
values of 16 clusters, when $\beta$ was inferred from X-ray observations (Host
e.t al.\ 2009). The black triangle is a value inferred for A1689 using galaxy
dynamics information (Lemze et al.\ 2011). 

\begin{figure*}
\centering
\makebox{
\epsfig{file=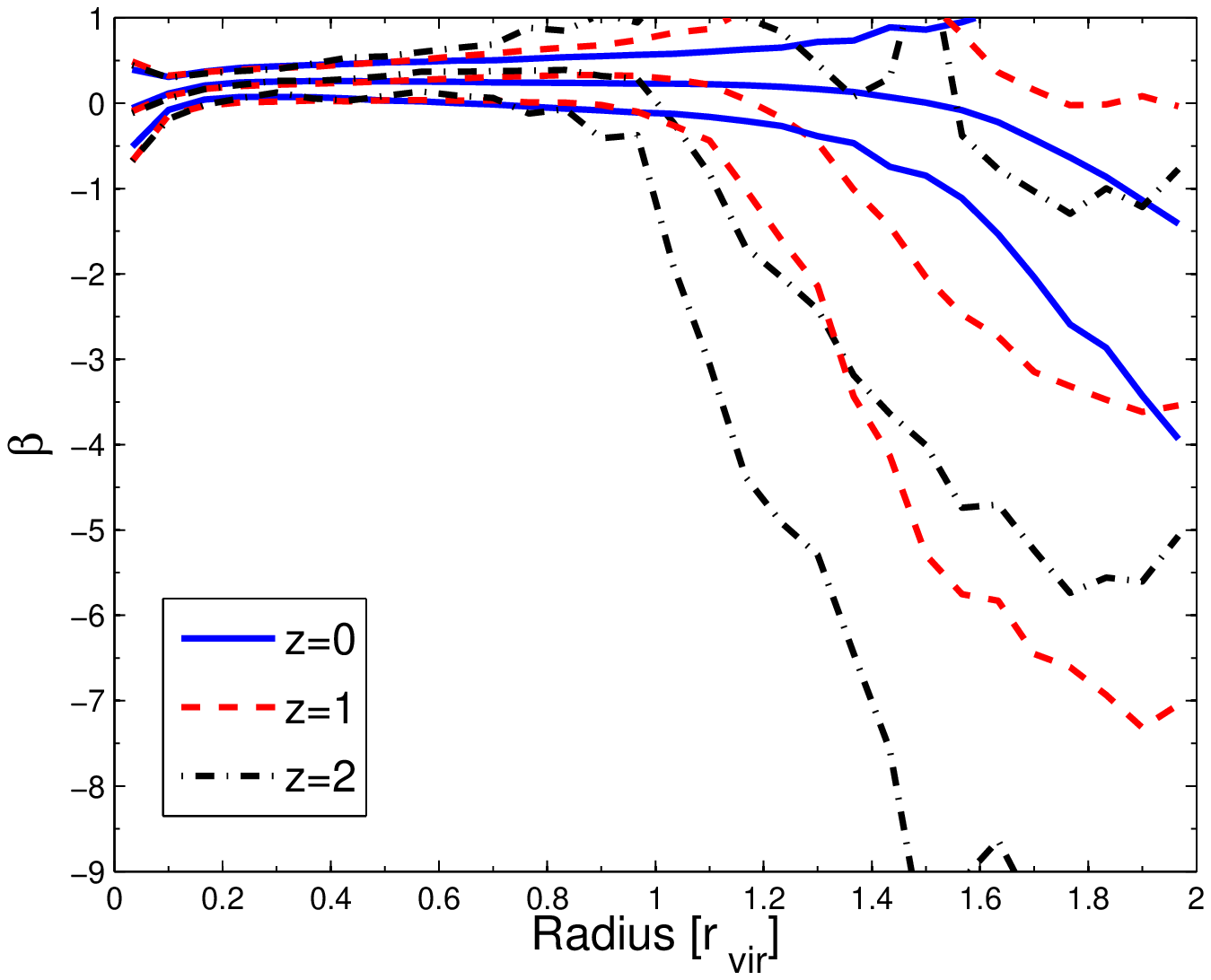, width=6cm,
clip=}
\epsfig{file=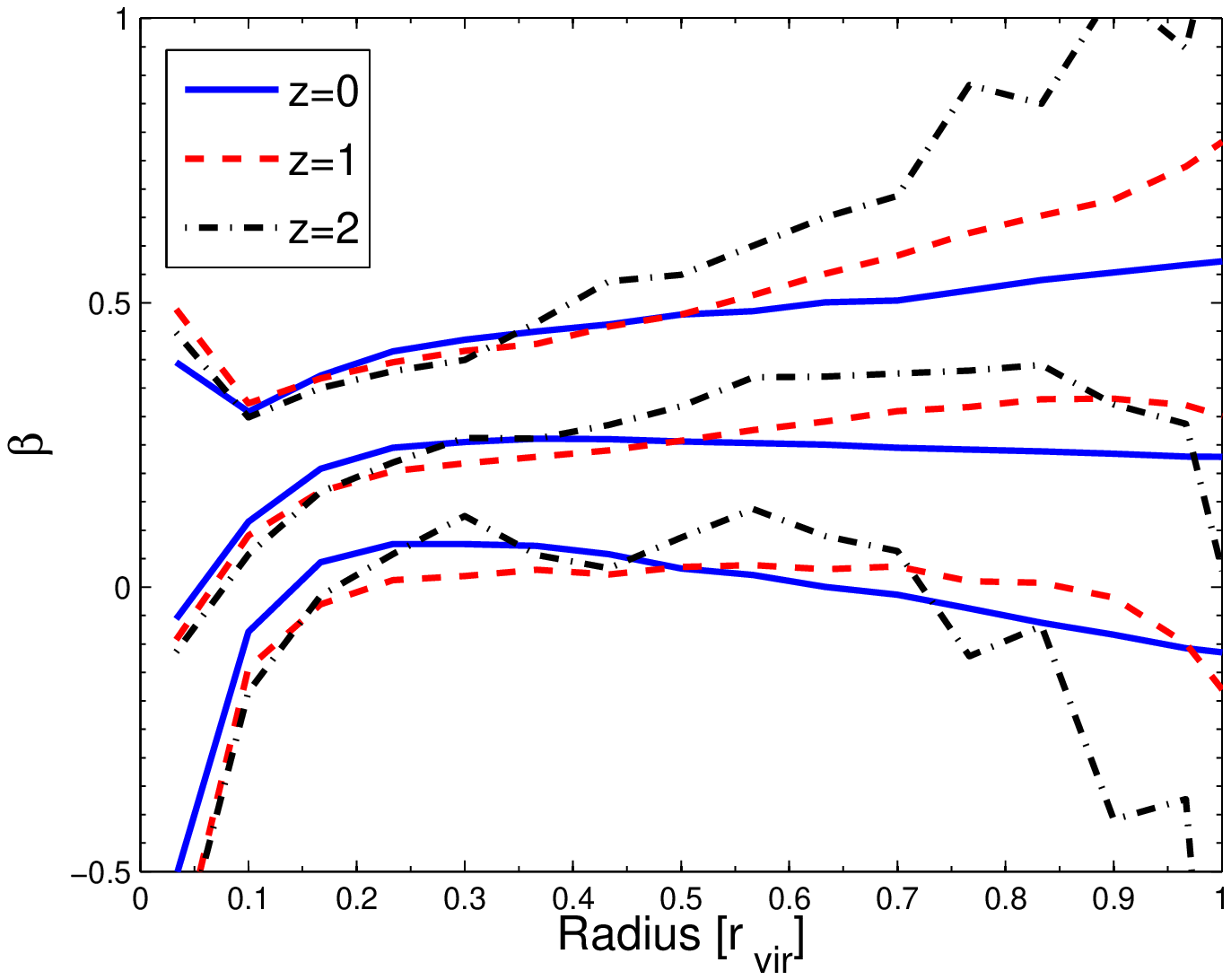, width=6cm, clip=}
\epsfig{file=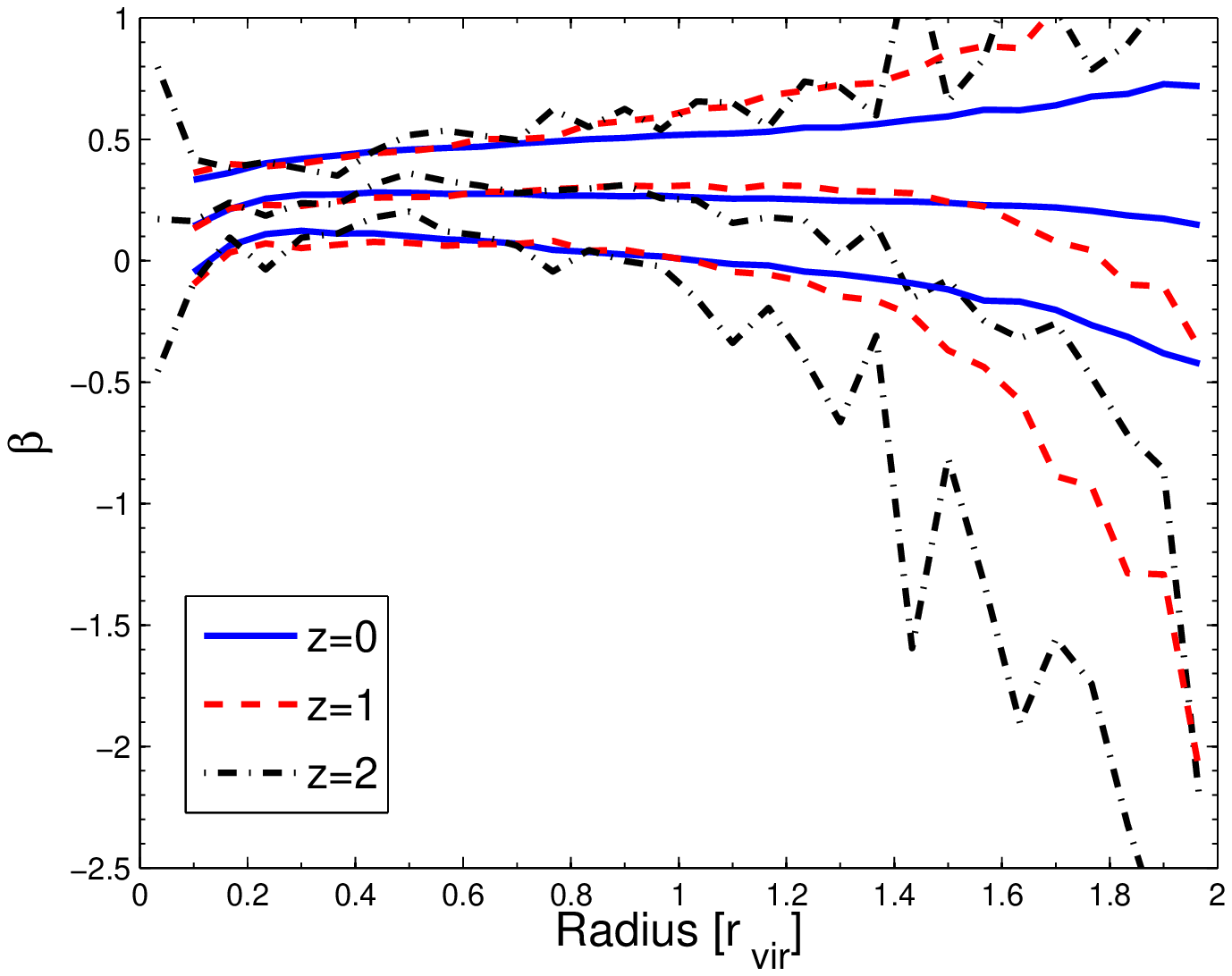, width=6cm, clip=}
}
\caption{Velocity anisotropy profiles at different redshifts. Left panel: using
spherical shells. Middle panel: the relatively
small variation of the anisotropy in the region $0-1$ $r_{\rm vir}$ in
spherical shells is shown separately. Right panel: using elliptical shells,
with elliptical radii, $r_{\rm e}$, scaled by $r_{\rm vir}$ of the spherical
radius. 
\label{beta_diff_redshift}}
\end{figure*}

\begin{figure}
\centering
\epsfig{file=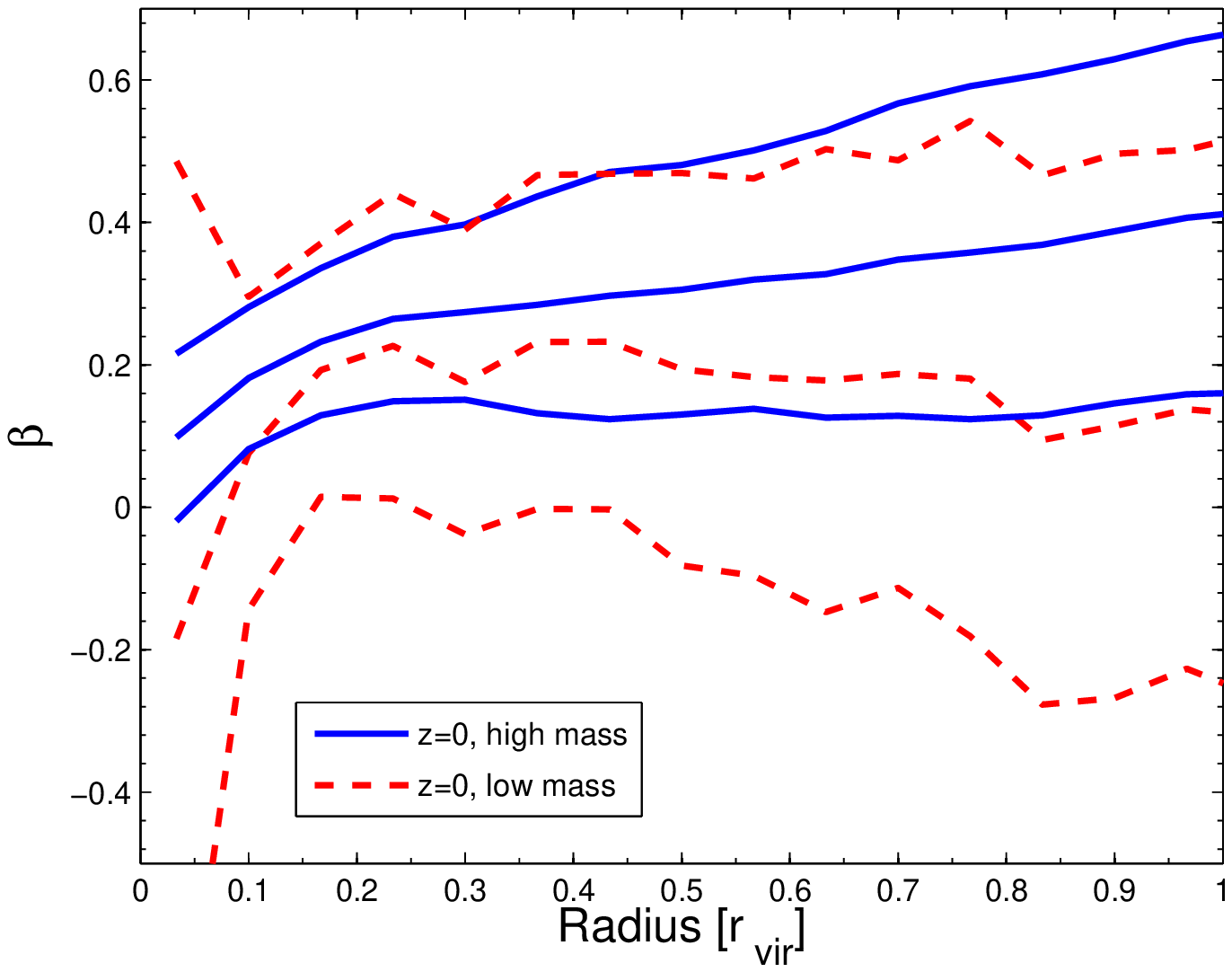, width=8cm, clip=}
\caption{Velocity anisotropy profiles of halos with different masses. 
Profiles of the 100 most (solid curve), $M\gtrsim 10^{15}$
h$_{0.7}^{-1}$ M$_{\odot}$, and least (dashed curve), $M \sim 10^{14}$
h$_{0.7}^{-1}$ M$_{\odot}$, massive halos at $z=0$ in spherical shells. 
\label{beta_diff_halo_mass}}
\end{figure}

Average DM velocity anisotropy profiles of all high-mass halos, $M_{\rm
vir}\geqslant 10^{14}$ h$^{-1}_{0.7}$ M$_{\odot}$, are plotted 
in figure~\ref{beta_diff_redshift} (including 1$\sigma$ uncertainty regions) 
for different redshifts and for spherical and elliptical shells. Using
spherical shells, the profiles are essentially similar at small radii, $r
\lesssim 0.3 r_{\rm vir}$, roughly independent of redshift. However, at larger
radii, $r \sim 0.7 r_{\rm vir}$, values of $\beta$ are somewhat higher in high
redshift halos. At even larger radii, $r\gtrsim r_{\rm vir}$, $\beta$ is lower
at higher redshifts. As the redshift increases, the scatter in $\beta$ increases
as a function of radius from $\sim 0.3 r_{\rm vir}$. Using elliptical shells,
the differences between the profiles decrease. 
In figure~\ref{beta_diff_halo_mass} we illustrate the DM velocity
anisotropy profiles for two mass ranges in spherical and elliptical shells,
respectively. The 100 most massive halos, $M\gtrsim 10^{15}$ h$_{0.7}^{-1}$
M$_{\odot}$, and
least massive halos, $M \sim 10^{14}$ h$_{0.7}^{-1}$ M$_{\odot}$, are compared
at redshift $z=0$.

As mentioned in \textsection~\ref{Criteria for relaxed clusters}, there are
various criteria for relaxed clusters. Here we use the $r_{\rm offset}$
criterion, since the difference between the relaxed and unrelaxed $\beta$
profiles is the largest (among the first three). We also mentioned in
\textsection~\ref{Criteria for relaxed clusters}, the threshold values of the
(quantitative) criteria according to which halos are classified as relaxed or
unrelaxed are quite arbitrary. However, in \textsection~\ref{appendix} we can
see that $r_{\rm offset}$ is more correlate with the halo ellipticity (which
relates to the relaxation level, see \textsection~\ref{Halos ellipticities at
different relaxation levels}) than the virial and corrected virial ratio. We
chose in our analysis to compare among the $\beta$ profiles by setting two
comparable numbers corresponding to the most and least relaxed phases.  
\begin{figure}
\centering
\epsfig{file=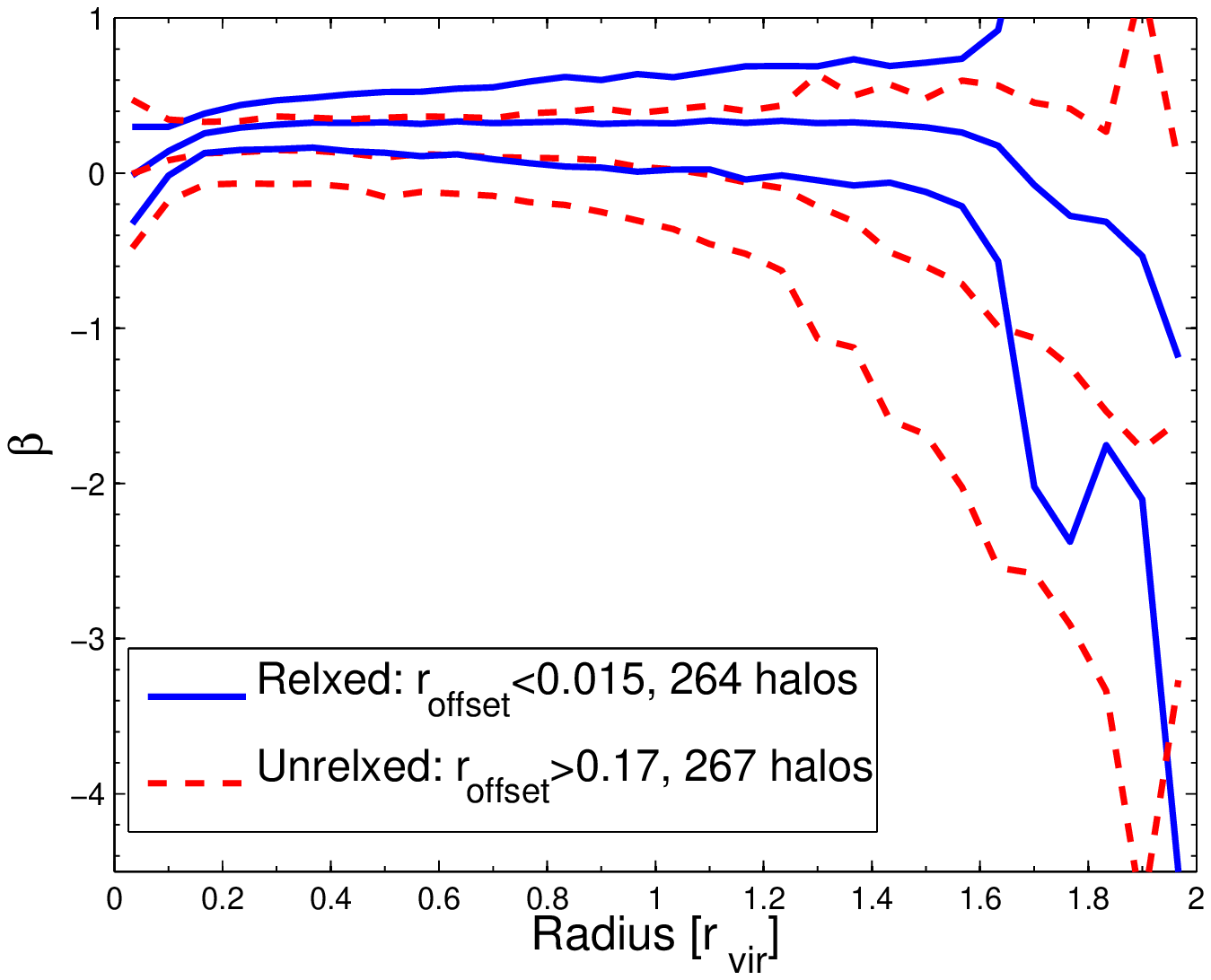,
width=8cm, clip=}
\caption{Velocity anisotropy profiles of relaxed (blue solid curves) and
unrelaxed (red dashed curves) halos analyzed in spherical shells. 
Relaxation gauged by the $r_{\rm offset}$ criterion: relaxed and unrelaxed
halos have $r_{\rm offset}<0.015$ (264 halos) and $r_{\rm offset}>0.17$ (267 
halos), respectively.
\label{beta_diff_relaxation_state_spheric}}
\end{figure} 
In figure~\ref{beta_diff_relaxation_state_spheric} we plot the velocity
anisotropy profile of relaxed versus unrelaxed halos using spherical shells 
when the distinction is made according to the $r_{\rm offset}$ criterion, where
relaxed and unrelaxed halos have $r_{\rm offset} < 0.015$ (264 halos) and
$r_{\rm offset} > 0.17$ (267 halos), respectively. Using the
virial ratio to distinguish between relaxed and unrelaxed halos yielded very 
similar $\beta$ profiles, and therefore these are not shown here. 
At radii smaller than the virial radius applying the $r_{\rm offset}$ 
criterion results in flattened velocity anisotropy profiles of the unrelaxed 
halos with respect to the relaxed halos.   

The $\beta$ profile of relaxed versus unrelaxed halos in elliptical
shells are not appreciably different than the ones using spherical shells.
\begin{figure}
\centering
\epsfig{file=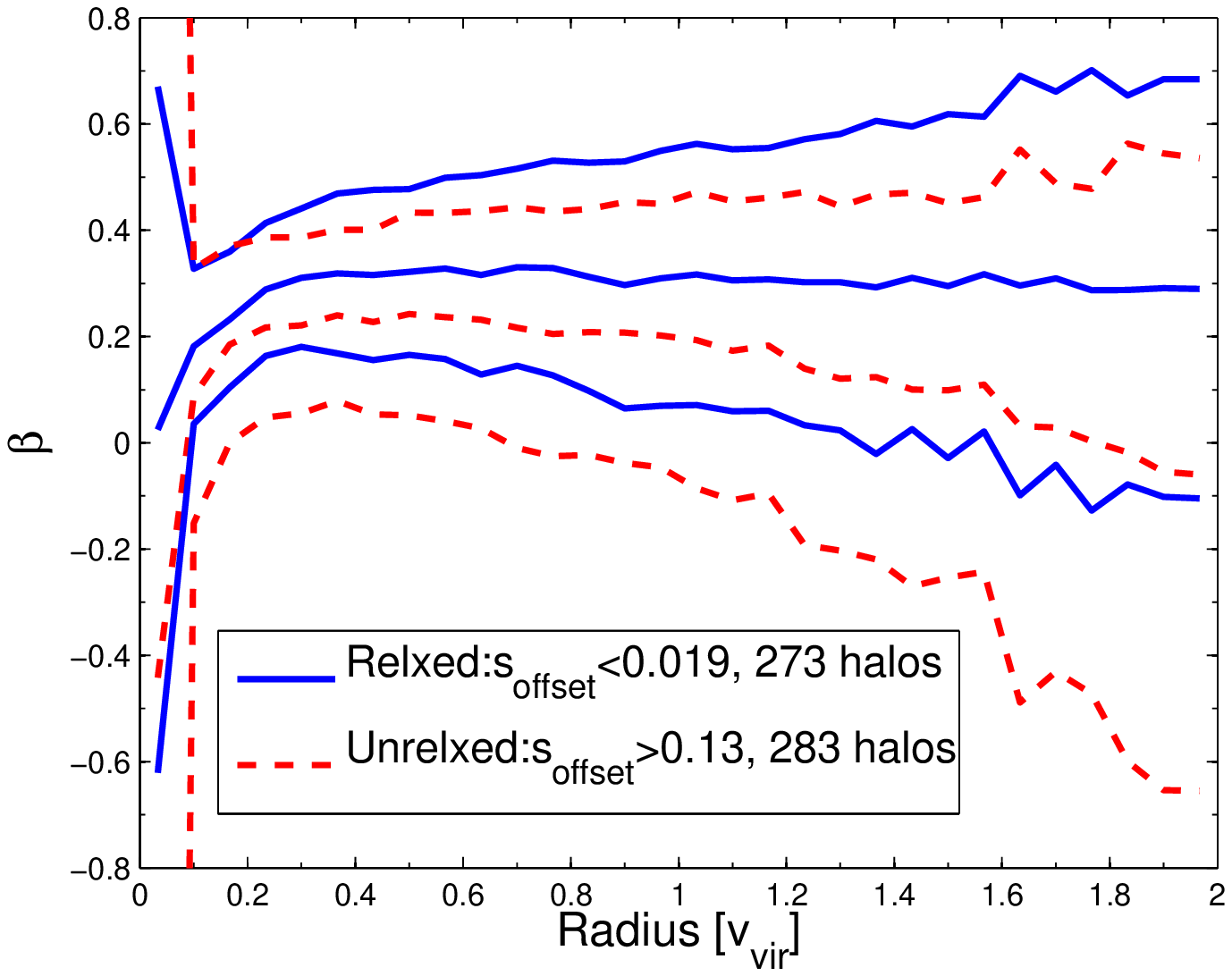,
width=8cm, clip=}
\caption{Velocity anisotropy profiles of relaxed (blue solid curves) and
unrelaxed (red dashed curves) halos in elliptical shells. Relaxation gauged by
the $r_{\rm offset}$ criterion: relaxed and unrelaxed halos have $s_{\rm
offset}<0.019$ (273 halos) and $r_{\rm offset}>0.13$ (283 halos), respectively.
\label{beta_diff_relaxation_state_elliptic}}
\end{figure} In figure~\ref{beta_diff_relaxation_state_elliptic} we plot the
velocity anisotropy profile of relaxed versus unrelaxed halos using elliptical
shells when the distinction is made according to the $r_{\rm offset}$ criterion.
The profiles are similar to those for spherical shells except for a smaller
decline and with a smaller scatter at large radii. Here we can also check if
the indication that $r_{\rm offset}$ is a better relaxation proxy depends
on the chosen threshold values. The correlations between all of the five
criteria and halos ellipticities are shown in \textsection~\ref{appendix}.
Out of the three relaxation criteria we focused ($r_{\rm offset}$, $2T/|U|$,
and $(2T-Es)/|U|$), the correlation between $r_{\rm offset}$ and the halos
ellipticities is the highest. Therefore, we conclude that our findings do
not depend on the threshold value.

\subsection{$\gamma$-$\beta$ ratio}

\begin{figure*}
\centering
\makebox{
\epsfig{file=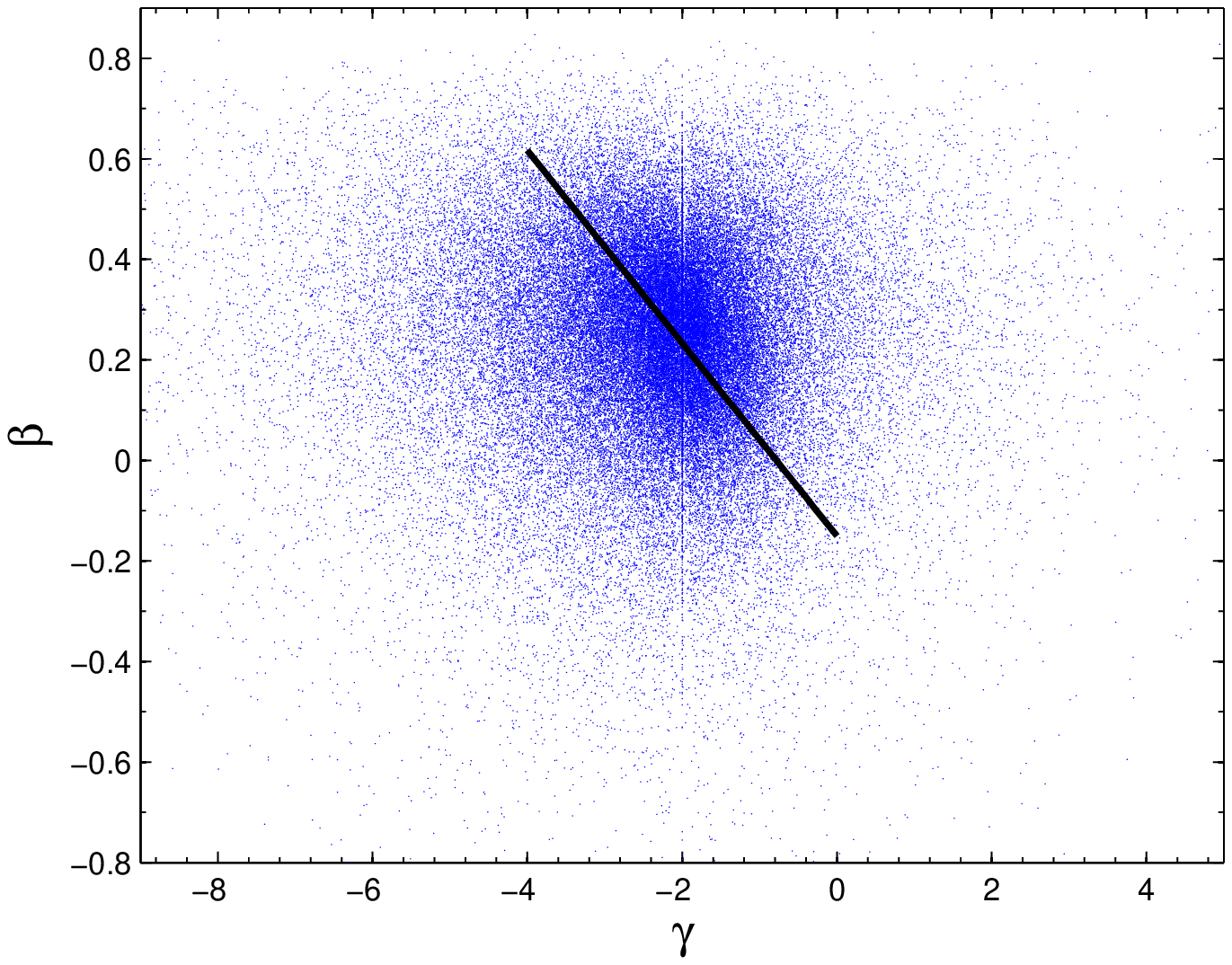, width=6cm, clip=}
\epsfig{file=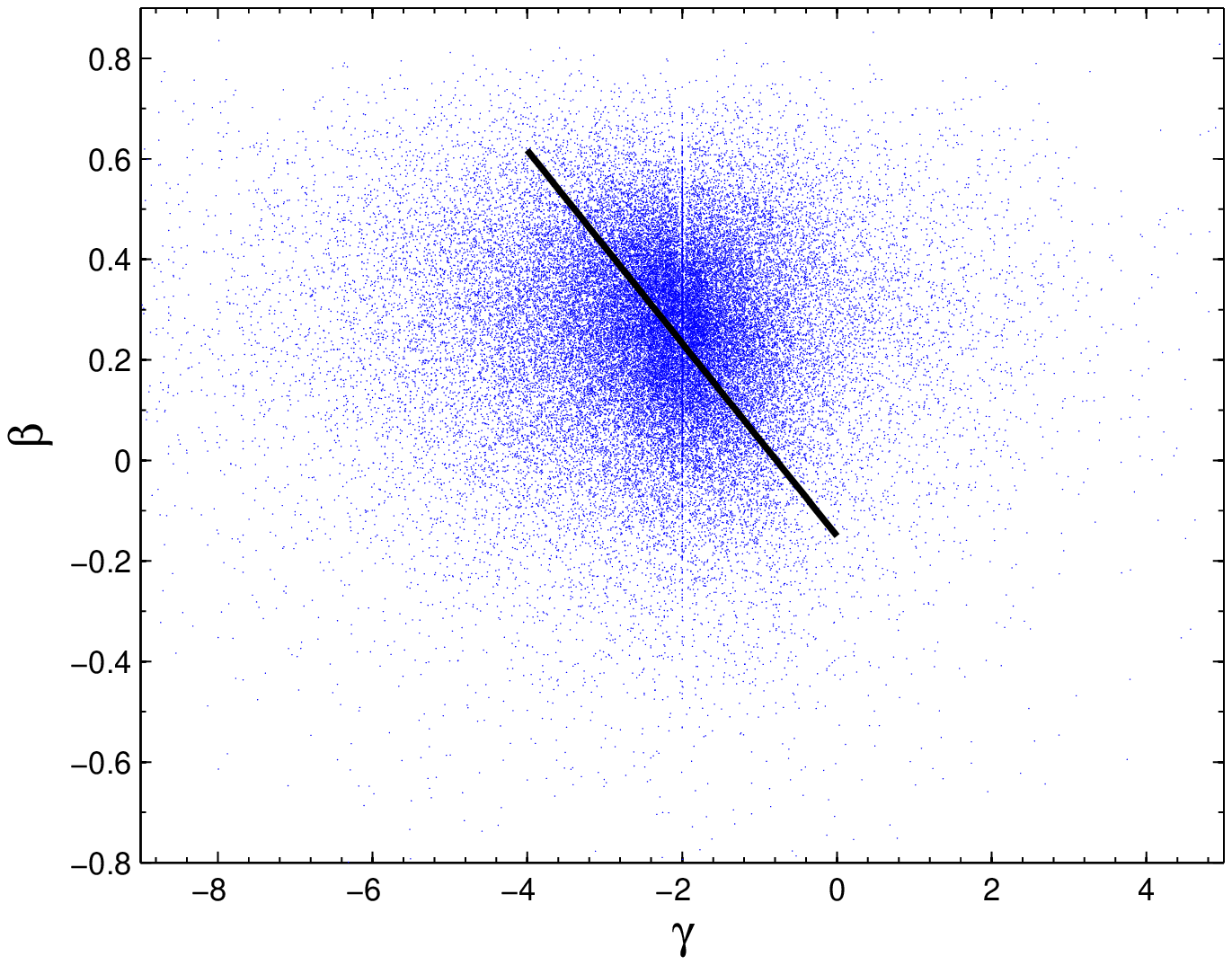, width=6cm, clip=}
\epsfig{file=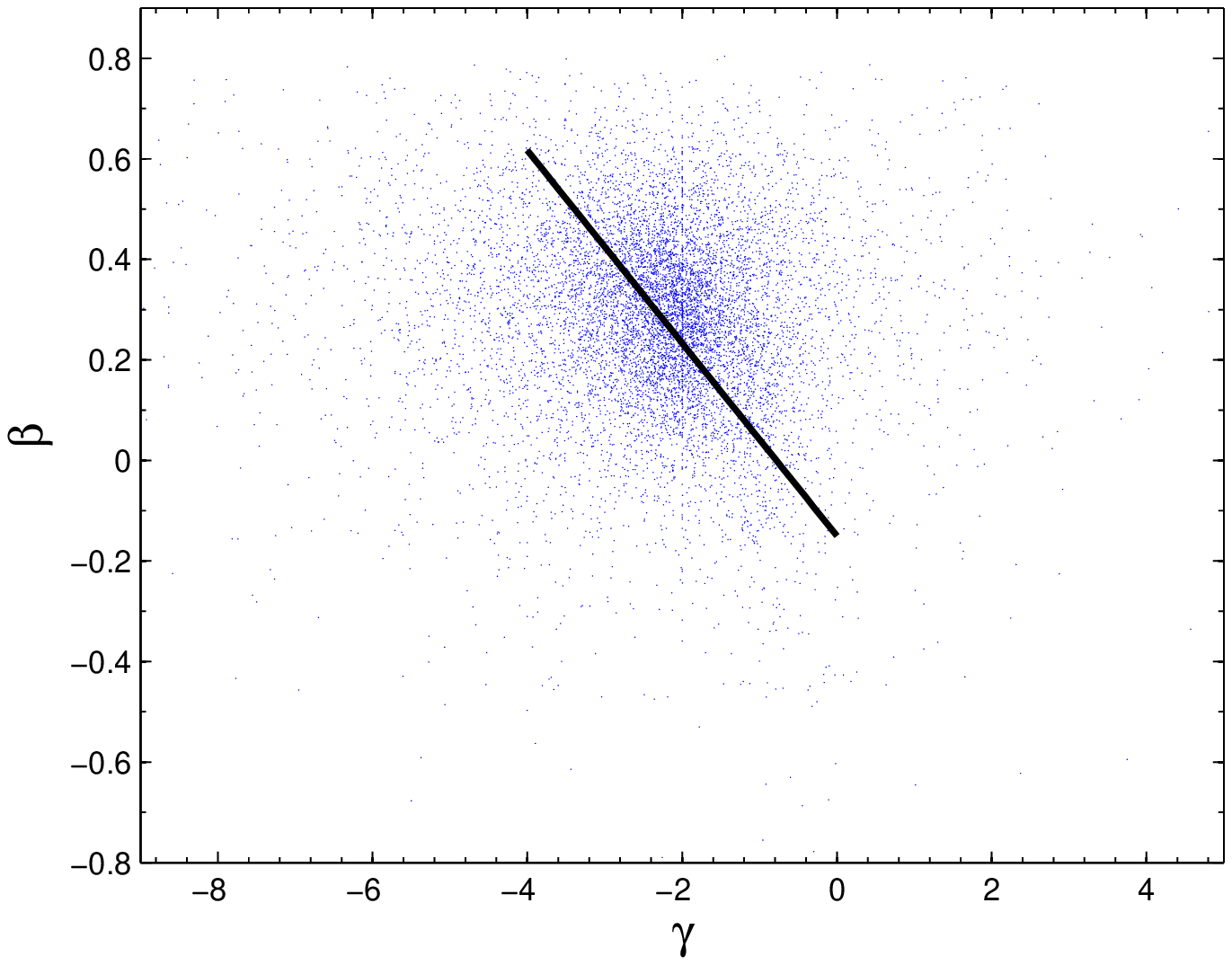, width=6cm, clip=}
}
\caption{Velocity anisotropy vs. radial density slope at $z=0$. Plotted are
values corresponding to all shells and halos (left panel), all shells of
relaxed halos according to the virial relation criterion $2T/|U|<1.35$
(middle panel), and all shells of highly relaxed halos, specified by
$2T/|U|<1.35$ and $r_{\rm offset}<0.025$ (right panel). The HM06 relation is
plotted in the $-4<\gamma<0$ range (black solid
curve).\label{beta_gamma_z0}}
\end{figure*}

\begin{figure}
\centering
\epsfig{file=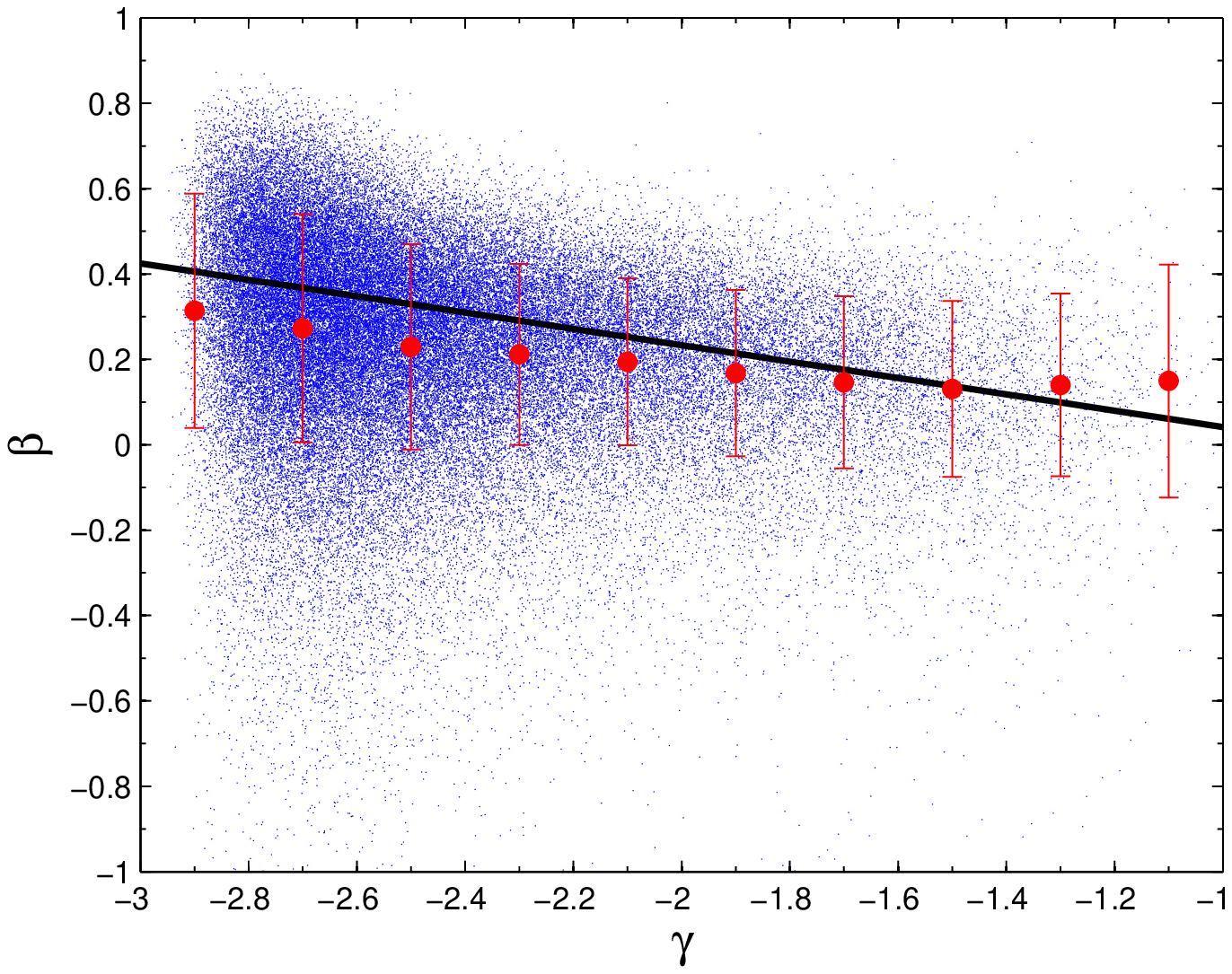, width=6cm, clip=}
\caption{Velocity anisotropy vs. radial density slope at $z=0$. Plotted are
values for all shells and halos, with $\gamma$ inferred from an NFW fit
(blue dots). Red circles with error bars show the median anisotropy profile and
$1\sigma$ dispersion. The HM06 relation is plotted in the $-4<\gamma<0$ range
(black solid curve).
\label{beta_gamma_NFW_z0}}
\end{figure}

\begin{figure}
\centering
\epsfig{file=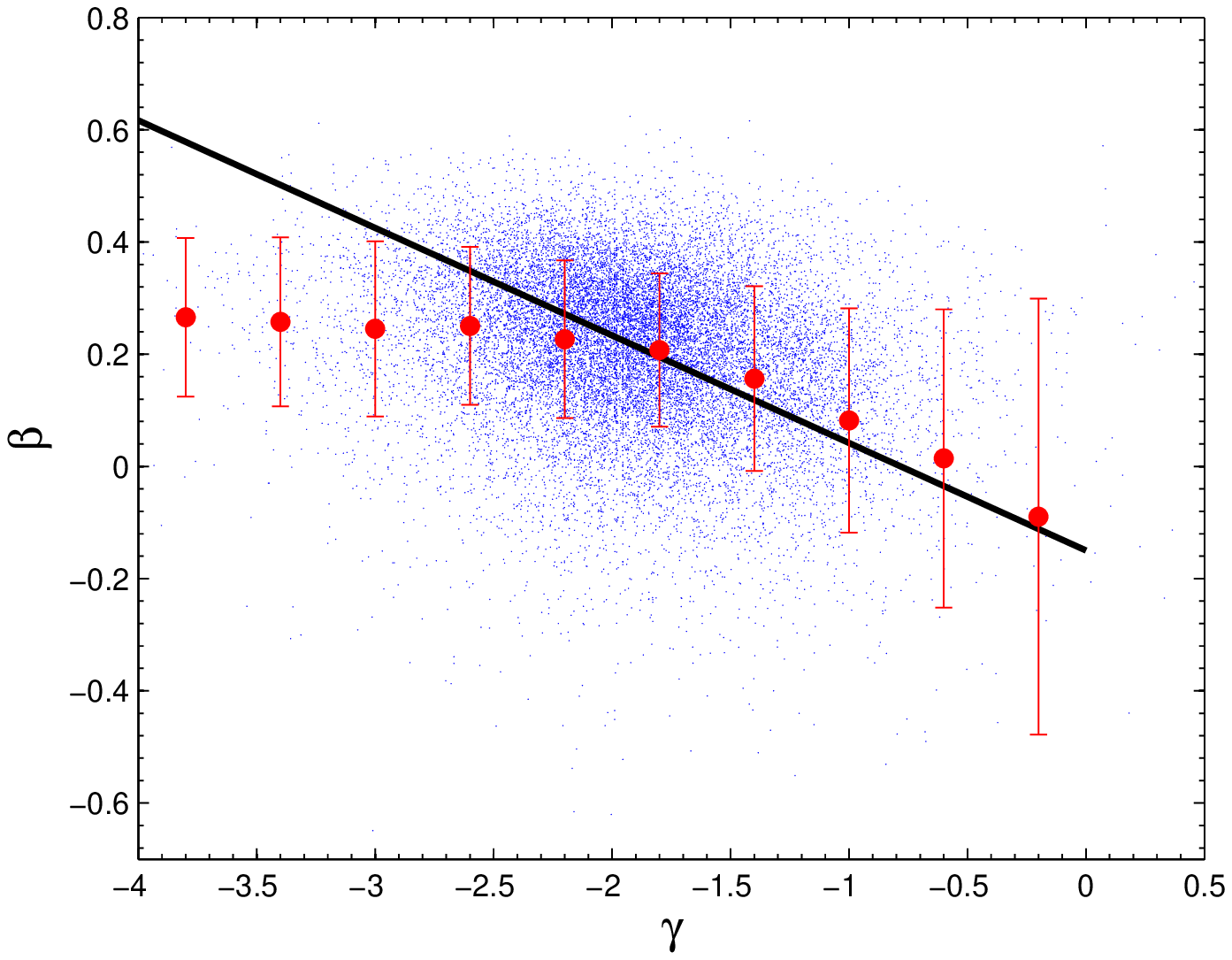, width=6cm, clip=}
\epsfig{file=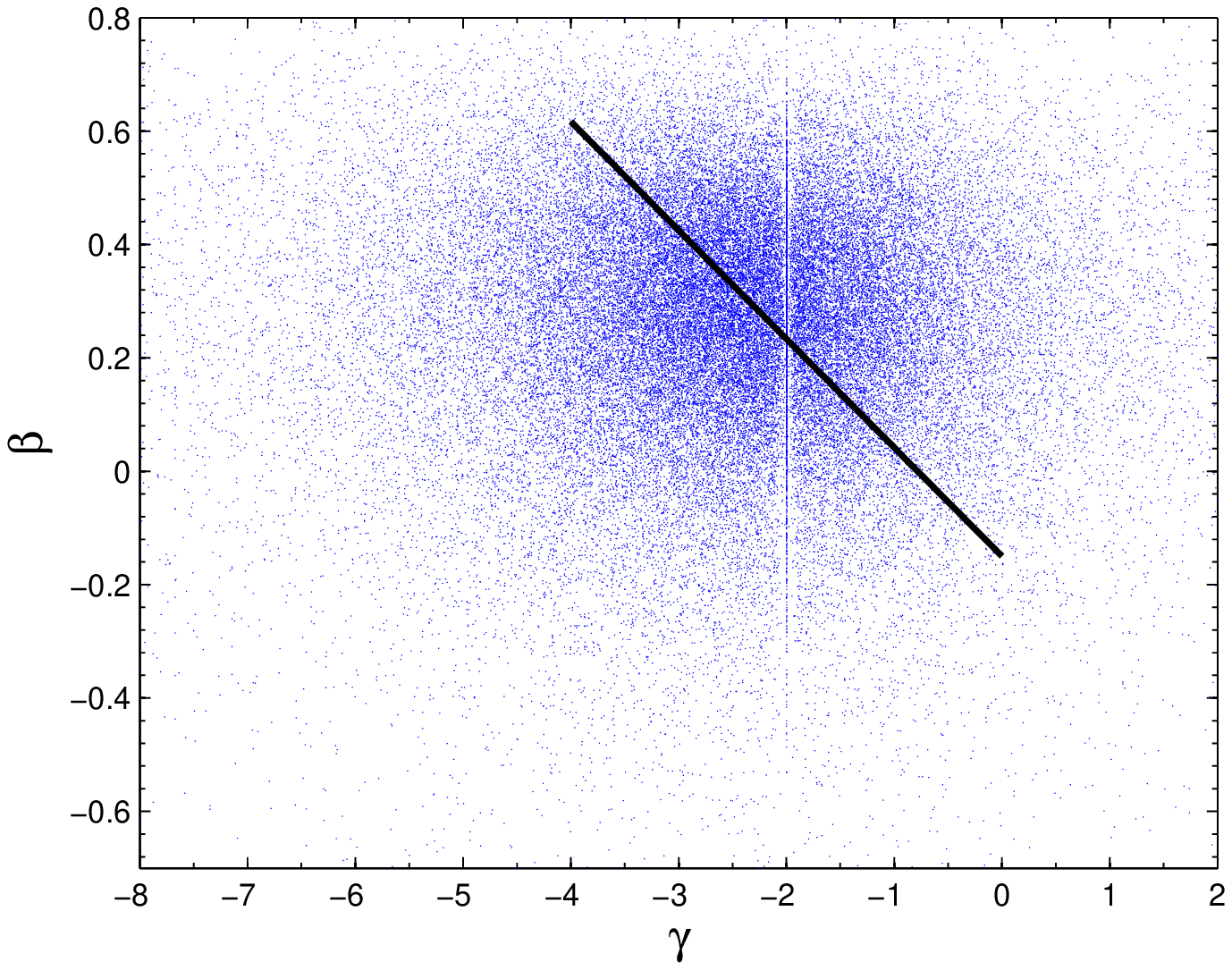, width=6cm, clip=}
\caption{Velocity anisotropy vs. radial density slope at $z=0$. Top panel:
plotted are values for the four inner shells, $0<r< 0.3 r_{\rm vir}$, and red
circles with error bars show the median anisotropy profile and $1\sigma$
dispersion. Bottom panel: values for the outer shells, $0.3 r_{\rm vir} <r<
r_{\rm vir}$. The HM06 relation is depicted in the $-4<\gamma<0$ range (black
solid curve). The vertical line at $\gamma \simeq -2$ is an artifact due to the
fact that $\gamma$ here is a discrete slope profile calculated between two
bins with a low number of particles. This artifact can also be seen in
figure~\ref{beta_gamma_z0}, though it is more prominent here.
\label{beta_gamma_z0_diff_shells}}
\end{figure}

As was mentioned in \textsection~\ref{Introduction}, the question of whether
$\gamma$ and $\beta$ are correlated is of both theoretical and
practical interest. In figure~\ref{beta_gamma_z0} we show the velocity
anisotropy vs. the radial density slope for all shells in all halos (left
panel), all shells of relaxed halos according to the virial relation criterion
$2T/|U|<1.35$ (middle panel), and all shells of highly relaxed halos with 
$2T/|U|<1.35$ and $r_{\rm offset}<0.025$ (right panel). For each halo we
checked the maximum grid level and determined the halo minimum spatial
resolution. Unresolved shells were not included in the analysis. The black
curve reproduces the Hansen \& Moore relation, $\beta(\gamma) = 1 -
1.15(1+\gamma/6)$, for the $-4<\gamma<0$ range. In
figure~\ref{beta_gamma_NFW_z0} we drew the same quantities for all shells in all
halos, assuming NFW-distributed density profiles. Red circles with error bars
show the median anisotropy profile and $1\sigma$ dispersion. Note that in this
plot the $\gamma$ range is $(\sim -2.9,-1)$, not ($-3$,$-1$), due to the finite
binned values of the radius.
Finally, figure~\ref{beta_gamma_z0_diff_shells} describes the velocity
anisotropy against the radial density slope of the four inner 
($0<r< 0.3 r_{\rm vir}$, top panel) and all the other 
($0.3 r_{\rm vir} <r< r_{\rm vir}$, bottom panel) shells. This boundary 
value was chosen since shells included within this radius, $0.3 r_{\rm vir}$,
display the strongest $\gamma$-$\beta$ correlation. In the panel showing values
for the inner shells we also plotted red circles with error bars that show the
median anisotropy profile and $1\sigma$ dispersion.

\begin{figure}
\centering
\epsfig{file=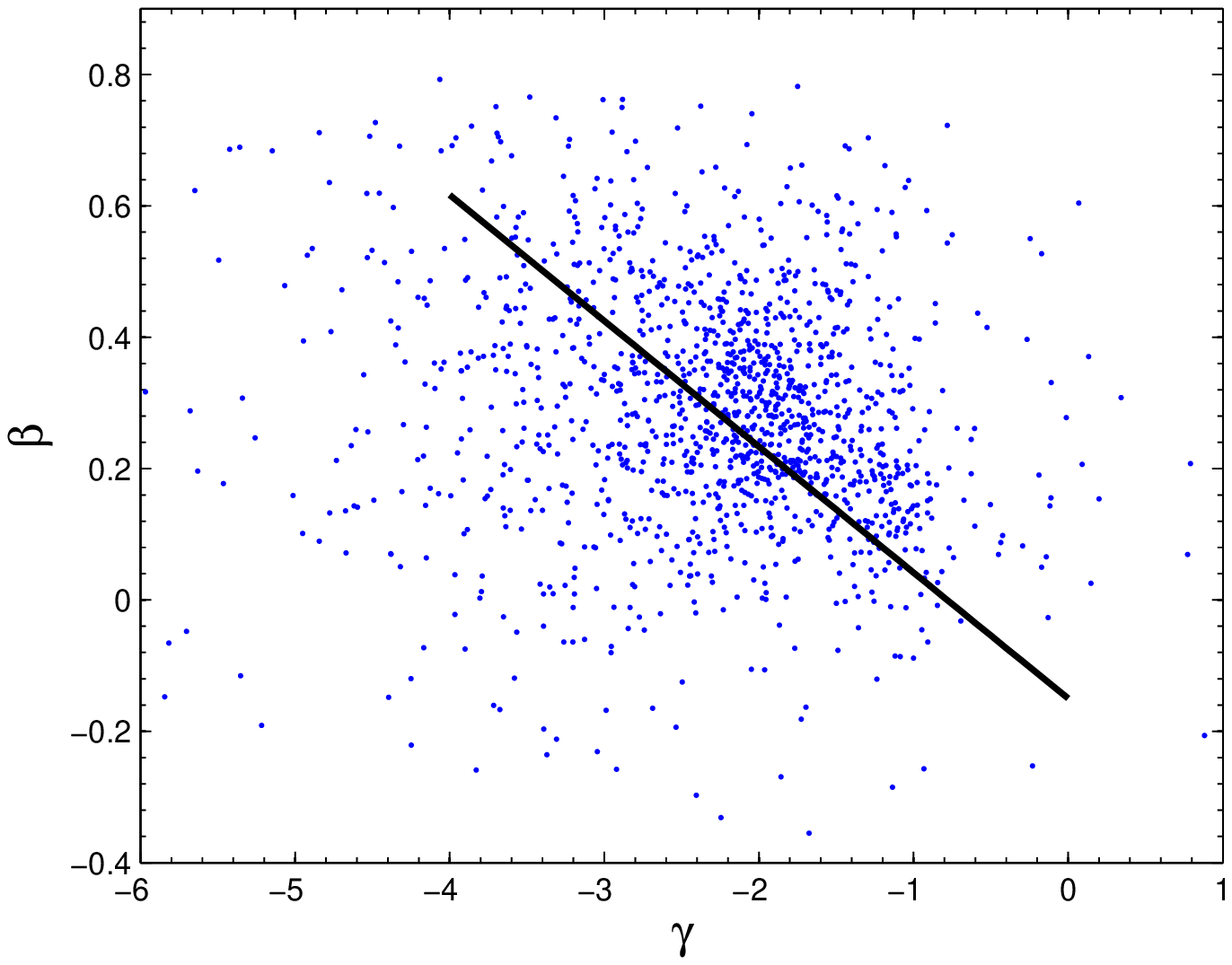, width=6cm, clip=}
\caption{Velocity anisotropy vs. radial density slope at $z=0$ for the 100 most
massive halos. 
\label{beta_gamma_z0_100mostmassive}}
\end{figure}

We have made two checks in order to see if our results are different when there
is a higher number of particles in each shell. First we decreased the number of
bins from 15 per virial radius to 5. The main effect was a decrease in the
$\gamma$ value range. This can be due to the fact that with a lower number of
bins the minimum and maximum bins are closer and the mean has lower variance.
In the second test we checked the 100 most massive halos, which obviously have 
higher number of particles per shell than the full sample. In both cases the
correlation between $\beta$ and $\gamma$ did not increase significantly, as can
be seen in figure~\ref{beta_gamma_z0_100mostmassive} for the $\beta$ and
$\gamma$ plot of the 100 most massive halos.

As expected, since the $\beta$ values were taken within the virial radius, and
since their largest difference between spherical and elliptical shells is at $r>
r_{\rm vir}$, when we repeated this above analysis with triaxial halos, the 
results were essentially the same as those obtained for spherical halos. 

\section{Discussion}
\label{Discussion}

Significant progress has recently been made in the ability to deduce the 
kinematic properties of DM in clusters from galaxy dynamics and X-ray
measurements. Comparison of these properties with results from numerical
simulations can clearly test analysis methods and add new insights on DM phase
space occupation. We presented results from an analysis of DM velocities in 
$6019$ halos with masses $M\geqslant 10^{14}$ h$_{0.7}^{-1}$ M$_{\odot}$ at 
redshift $z=0$, drawn from one of the largest ever hydrodynamic cosmological 
AMR simulations.

We found that each halo ellipticity depends strongly on the major axis length.
This is because for each halo ellipsoids with different sizes include different
parts of substructure and infalling clumps, which obviously affect the
ellipticity, if they are inside or close to the ellipse. However, averaging over
many halos, $\sim 10^3$, the halo ellipticity first decreases a small amount,
until $\sim (1.5-2)r_{\rm vir} $, then the averaged ellipticity increases. In
other words on average the halos are more elliptical at small radii then with
increasing radius they become more spherical, and then elliptical again, since
at large radii other structures are located inside the ellipsoid and this on
average increases the ellipticity. However, the change is small, even in the
relaxed sample, and over the radius range $(0.5-3)r_{\rm vir}$ the axis ratios
are essentially constant, $<q>\approx 0.66$ and $<s>\approx 0.5$ and $<q>\approx
0.7$ and $<s>\approx 0.54$, for the whole $\sim 10^3$ halo sample and for the
relaxed one, respectively, especially compared to the large scatter. This
ellipticity vs radius behavior is in agreement with Allgood et al.\ (2006) who
found that halos become more spherical up to $r=r_{\rm vir}$, and Bailin \&
Steinmetz (2005), who found the ellipticity is quite constant compared with the
scatter. Kuhlen et al.\ (2007), Hayashi et al.\ (2007), and Vera-Ciro et al.\
(2011) used various N-body simulations and found that, similarly to cluster-size
halos, in galaxy-size halos the core is more elliptical than a few hundreds of
kpc away. However, Hopkins, Bahcall, \& Bode\ (2005) found that cluster
cores are less elliptical than cluster outskirts

We have employed five different relaxation criteria and focused on three of
them, $r_{\rm offset}$, $2T/|U|$, and $(2T-Es)/|U|$; see the Appendix 
for the correlations of these three with the two other relaxation proxies we
also considered. Relaxed  halos tend to be more spherical while unrelaxed halos
tend to be more prolate. This is seen using various relaxation criteria, e.g. 
$r_{\rm offset}$, $2T/|U|$, $(2T-Es)/|U|$. This can be explained by the
evolution of clusters from highly aligned and elongated systems at early times
to lower alignment and elongation at present, which reflects the hierarchical
and filamentary nature of structure formation (Hopkins, Bahcall, \& Bode\ 2005).
Indeed, Vera-Ciro et al.\ (2011), who analyzed Aquarius data, found that 
$q$ increases with time. This is also in agreement with Shaw et al.\ (2006) 
who found that low mass halos, which are older than their higher mass 
counterparts and therefore had a longer time to relax dynamically, attain a 
more spherical morphology than high mass halos. 

Our study indicates that the profiles of cluster DM velocity anisotropy
have a similar pattern when these are averaged over all halo masses, redshifts,
and relaxation stages, even though there is a considerable scatter in magnitude
due to the large differences in the $\beta$ profile of individual halos. A
typical behavior is a rising $\beta$ profile from a nearly vanishing central
value, leveling off at $r\sim 0.2 r_{\rm vir}$, out to large radii of at 
least $(1.5-2)r_{\rm vir}$. This plateau is clearer in elliptical shells and at
lower redshifts. The rising value from zero at small radii behavior is in
agreement with previous works (e.g. Crone et al.\ 1994; Tormen et al.\ 1997;
Thomas et al.\ 1998; Eke et al.\ 1998;Colin et al.\ 2000; Rasia et al.\
2004). Interestingly, Lau, Nagai, \& Kravtsov\ (2010) showed that
the inclusion of radiative cooling and star formation in the simulation 
slightly lowers the $\beta$ values at $z=0$ but not at $z=1$. Unrelaxed 
halos have lower $\beta$ values, which indicates that the DM particles 
infall is less radial than in relaxed halos. Lower mass halos have, on 
average, lower $\beta$ levels at $r=r_{\rm vir}$, and therefore even 
shallower profiles. This behavior could possibly be due to the ability to 
reach higher accretion velocities at lower redshifts when the background 
density is much lower. However, our mass range, especially in elliptical 
shells, is very small, so these findings are preliminary. 

Some effort is now devoted to determine $\beta$ either by using the gas
temperature as a tracer of it, a method which was applied for 16 clusters at
intermediate radii (Host et al.\ 2009), or by examining galaxy velocities, as
has recently been demonstrated in the analysis of A1689 measurements (Lemze et
al.\ 2011). Interestingly, these indirect measurements of $\beta$ fall on the
upper end of the theoretically expected range. Both methods assume steady state,
i.e. $\overline{v}_r=0$. Clusters on the other hand are not in steady state,
since matter is continuously falling onto them. However, at small radii (the
indirect measurements radii range), i.e. $r \lesssim 0.7$ $r_{\rm vir}$, this
effect on $\beta$ is negligible, $\lesssim 4 \%$. In a recent work, Peter,
Moody, \& Kamionkowski (2010) used simulations and showed that for galaxy-size
halos, the expected $\beta$ increases with radius till about the virial radius
when DM particles decay to a slightly less massive particle with a large
decaying time, $\tau \gtrsim$ a few Gyr, and a high velocity kick, $v_{\rm kick}
\gtrsim 150$ km s$^{-1}$. 

Lastly, we find that there is some correlation between $\gamma$ and
$\beta$ at low radii, $r<0.3 r_{\rm vir}$, and that such a correlation
can be induced at all radii merely by assuming a prescribed DM density profile
(since when assuming a profile the $\gamma$ values are correlated and therefore
the scatter along the $\gamma$ axis is reduced).
The level of $\gamma$ -- $\beta$ correlation is very low at
large radii, $r>0.3 r_{\rm vir}$, even for very relaxed halos. Repeating the
same analysis with elliptical shells led to the same result. Indeed, most of 
the works that try to explain the relation between $\gamma$ and $\beta$ 
focus on the inner regions. Hansen et al.\ (2005) and Hansen (2009) assumed 
spherically symmetric systems in hydrostatic equilibrium, and that high 
energy tail of the tangential velocity distribution function follows that of 
the radial velocity distribution function, based on which they explain 
the $\gamma$-$\beta$ relation mainly at small radii where both $\gamma$ and 
$\beta$ are about zero. An \& Evans (2006, and references therein) 
have deduced that $\gamma \leq \beta$. Ciotti \& Morganti (2010) claimed 
that this inequality holds not only at the center, but also at larger radii 
in a very large class of spherical systems whenever the phase-space 
distribution function is positive.

\section{Conclusions}
\label{Conclusions}

Our main conclusions are as follows:
\begin{itemize}

\item Galaxy clusters are generally triaxial.

\item Relaxed halos tend to be more spherical while unrelaxed halos tend to 
be more prolate.

\item Low mass halos tend to be more relaxed (see \textsection~\ref{appendix}),
though this result is based on a narrow mass range and therefore should
be considered preliminary.

\item The ellipticity of each halo strongly depends on the semi-major axis
amplitude. However, averaging over many halos, $\sim 10^3$, there is no
significant difference in the ellipticity at ellipses with different semi-major
axis amplitudes, even in the relaxed sample.

\item The $r_{\rm offset}$ is a better relaxation proxy than $2T/|U|$ and
$(2T-Es)/|U|$ in the sense that its correlation with the halo ellipticity is
stronger.

\item The pressure term, $Es$, is larger in elliptical sells and in less
relaxed halos (when the relaxation criterion is $2T/|U|$).

\item DM velocity anisotropy profiles have a similar pattern when these are
averaged over all halo masses, redshifts, triaxiality, and 
relaxation stages, even though there is a considerable scatter in
magnitude due to the large differences in the $\beta$ profile of individual
halos. A typical behavior is a rising $\beta$ profile from a nearly vanishing
central value, leveling off at $r\sim 0.2 r_{\rm vir}$, out to large radii 
of at least $(1.5-2)r_{\rm vir}$. This plateau is clearer in elliptical 
shells and at lower redshifts.

\item DM velocity anisotropy indirect measurements fall on the upper end of the
theoretically predicted range.

\item There is some correlation between $\gamma$ and $\beta$ at low radii,
$r<0.3 r_{\rm vir}$, and that such a correlation can be induced at all radii
merely by assuming a prescribed DM density profile. The level of $\gamma$ --
$\beta$ correlation is very low at large radii, $r>0.3 r_{\rm vir}$, even for
very relaxed halos. 

\end{itemize}

The average expected DM velocity anisotropy has a similar pattern for all halo
masses, redshifts, triaxiality, and relaxation stages. DM velocity anisotropy 
indirect measurements fall on the upper edge of the theoretical expectations
(see figure~\ref{Different shells}). Though measured indirectly, the
estimations are derived by using two different surrogate measurement, i.e. 
X-ray and galaxy dynamics. So far the DM velocity anisotropy estimates 
were based on a very low number of clusters (16 via X-ray and 1 via 
galaxy dynamics). It will be important to contrast the theoretically predicted
values with results from a larger cluster sample. We plan to do this in CLASH.

\section*{ACKNOWLEDGMENT}
\noindent We thank Brain O'shea and Steen Hansen for helpful discussions. We
thank Steen Hansen also for very useful communication during the revision
process. We also thank Greg Bryan and the referee for very helpful comments.
This research is supported in part by NASA grant HST-GO-12065.01-A. Work at Tel
Aviv University is supported by US-IL Binational Science foundation grant 
2008452. The simulations were performed on the DataStar system at the San
Diego Supercomputer Center using LRAC allocation TG-MCA98N020.

\appendix

\section{The correlations between the axes ratios, relaxation proxies, and
$\beta$}
\label{appendix}

Here we present the correlations between the axes ratios $q$, $s$,
relaxation proxies $r_{\rm offset}$, $r_{\rm sub}$, $r_{\rm dp}$, $2T/|U|$,
$(2T-Es)/|U|$ (when $Es$ is estimated by the outer 20\% and 10\% shell volume
for comparison). In tables~\ref{Correlations table - all halos} and
\ref{Correlations table - q>=0.4} the correlations are calculated out of all
the halos and the ones with $q \geqslant 0.4$, respectively.

\begin{table}
\caption{Correlations Matrix - All halos \label{Correlations table - all
halos}}
\begin{center}
\begin{tabular}{|c||c|c|c|c|c|c|c|c|c|}
\hline
         &    $M_{\rm vir}$ & $q$    & $s$  & $r_{\rm offset}$ & $r_{\rm sub}$
& $r_{\rm dp}$ & $2T/|U|$ & $(2T-E_{s,20\%})/|U|$,20\% & $(2T-Es)/|U|$,10\%  \\
\hline
\hline
$M_{\rm vir}$ & 1 & -0.26 & -0.24 & 0.15 & 0.14 &
0 & 0.28 & 0.11 & 0.1\\     
\hline
$q$  & -0.26 & 1 & 0.75 & -0.32 & -0.33 & -0.08 &
-0.21 & -0.16 & -0.16\\             
\hline
$s$ & -0.24 & 0.75 & 1 & -0.38 & -0.39 & -0.1 &
-0.23 & -0.17 & -0.17\\              
\hline
$r_{\rm offset}$  & 0.15 & -0.32 & -0.38 & 1 & 0.86 &
0.21 & 0.38 & -0.09 & -0.09\\ 
\hline
$r_{\rm sub}$ & 0.14 & -0.33 & -0.39 & 0.86 & 1 &
0.28 & 0.38 & -0.06 & -0.06\\     
\hline
$r_{\rm dp}$ & 0 & -0.08 & -0.1 & 0.21 & 0.28 &
1 & 0.09 & 0.1 & 0.09\\        
\hline
$2T/|U|$ & 0.28 & -0.21 & -0.23 & 0.38 & 0.38 & 0.09 &
1 & 0.41 & 0.31\\           
\hline
$(2T-Es)/|U|$,20\% & 0.11 & -0.16 & -0.17 & -0.09 & -0.06 &
0.1 & 0.41 & 1 & 0.96\\ 
\hline
$(2T-Es)/|U|$,10\% & 0.1 & -0.16 & -0.17 & -0.09 & -0.06 &
0.09 & 0.31 & 0.96 & 1\\    
\hline

\end{tabular}
\end{center}
\end{table}

\begin{table}
\caption{Correlations Matrix - Halos with $q \geqslant 0.4$ \label{Correlations
table - q>=0.4}}
\begin{center}
\begin{tabular}{|c||c|c|c|c|c|c|c|c|c|c|}
\hline
         &  $M_{\rm vir}$   &  $q$    & $s$  & $r_{\rm offset}$ & $r_{\rm
sub}$ & $r_{\rm dp}$ & $2T/|U|$ & $(2T-Es)/|U|$,20\% & $(2T-Es)/|U|$,10\%    \\
\hline
\hline
$M_{\rm vir}$  & 1 & -0.23 & -0.21 & 0.13 & 0.1 &
-0.05 & 0.26 & 0.09 & 0.08\\    
\hline
$q$   & -0.23 & 1 & 0.71 & -0.24 & -0.23 & 0.02 &
-0.15 & -0.11 & -0.12\\             
\hline
$s$  & -0.21 & 0.71 & 1 & -0.32 & -0.31 & -0.03 &
-0.16 & -0.12 & -0.13\\              
\hline
$r_{\rm offset}$ & 0.13 & -0.24 & -0.32 & 1 & 0.85 &
0.22 & 0.35 & -0.14 & -0.14\\  
\hline
$r_{\rm sub}$ & 0.1 & -0.23 & -0.31 & 0.85 & 1 &
0.22 & 0.33 & -0.14 & -0.13\\     
\hline
$r_{\rm dp}$ & -0.05 & 0.02 & -0.03 & 0.22 & 0.22 & 1
& 0.02 & 0.04 & 0.04\\      
\hline
$2T/|U|$ & 0.26 & -0.15 & -0.16 & 0.35 & 0.33 & 0.02 &
1 & 0.38 & 0.27\\        
\hline
$(2T-Es)/|U|$,20\% & 0.09 & -0.11 & -0.12 & -0.14 & -0.14 &
0.04 & 0.38 & 1 & 0.96\\  
\hline
$(2T-Es)/|U|$,10\% & 0.08 & -0.12 & -0.13 & -0.14 & -0.13 &
0.04 & 0.27 & 0.96 & 1\\  
\hline

\end{tabular}
\end{center}
\end{table}

\end{document}